\documentclass[sn-mathphys-num]{sn-jnl} 

\newcommand{\dd}{\,\mathrm{d}}
\newcommand{\tr}{\,\mathrm{tr}}
\newcommand{\ee}{\,\mathrm{e}}

\usepackage{amsmath}
\usepackage{amssymb}
\usepackage{mathtools}
\usepackage{hhline}
\usepackage{multirow, makecell}
\usepackage{physics}
\usepackage{esint}
\usepackage[titles]{tocloft}
\usepackage{bm}
\usepackage{makecell}
\usepackage{relsize}
\usepackage{graphicx}
\usepackage{algorithm}
\usepackage{algpseudocode}
\usepackage{booktabs}
\usepackage{tocloft}
\usepackage{hyperref}
\usepackage{eqparbox}

\theoremstyle{thmstyleone} 
\newtheorem{lemma}{Lemma}
\newtheorem{proposition}{Proposition}
\newtheorem{theorem}{Theorem}
\newtheorem{corollary}{Corollary}
\newtheorem*{example*}{Example}
\newtheorem*{remark*}{Remark}
\newtheorem*{remarks*}{Remarks}

\makeatother
\makeatletter
\newcommand{\mathleft}{\@fleqntrue\@mathmargin0pt}
\newcommand{\mathcenter}{\@fleqnfalse}

\def\plus{\text{\smaller +}}
\def\minus{\text{--}}

\newcommand{\ra}[1]{\renewcommand{\arraystretch}{#1}}

\begin{document}
\jyear{2025}
\title[Cumulant Structures]{Cumulant Structures of Entanglement Entropy}

\author{\fnm{Youyi} \sur{Huang}} 
\author{\fnm{Lu} \sur{Wei}} 

\affil{\orgname{\center Department of Computer Science \\ Texas Tech University} \\ \orgaddress{\city{Lubbock}, \state{Texas} \postcode{79409}, \country{USA}}}

\abstract{We present a new method to derive exact cumulant expressions of any order of von Neumann entropy over Hilbert-Schmidt ensemble. The new method uncovers hidden cumulant structures that decouple each cumulant in a summation-free manner into its lower-order joint cumulants involving families of ancillary statistics. Importantly, the new method is able to avoid the seemingly inevitable task of simplifying nested summations of increasing difficulty that prevents the existing method in the literature to obtain higher-order cumulants.}


\maketitle
{\pagestyle{plain}\tableofcontents}

\section{Introduction}
\subsection{Cumulants of entanglement entropy}\label{sec1.1}
Consider a composite quantum system that consists of two subsystems $A$ and $B$ of Hilbert space dimensions $m$ and $n$, respectively. Without loss of generality, we assume $m\leq n$. The Hilbert space of this bipartite system equals the tensor product of Hilbert spaces of the subsystems, $\mathcal{H}_{A}\otimes\mathcal{H}_{B}$. A random pure state of the composite system is represented by a linear combination of the random coefficients $c_{i,j}$ and the complete bases $\left\{\ket{i^{A}}\right\}$ of $\mathcal{H}_{A}$ and $\left\{\ket{j^{B}}\right\}$ of $\mathcal{H}_{B}$ as $\ket{\psi}=\sum_{i=1}^{m}\sum_{j=1}^{n}c_{i,j}\ket{i^{A}}\otimes\ket{j^{B}}$. The reduced density matrix $\rho_{A}=\tr_{B}(\rho)$ of the smaller subsystem $A$ is obtained as the partial trace of the full density matrix $\rho=\ket{\psi}\bra{\psi}$ over the other subsystem $B$. The eigenvalues $\lambda_{i}\in[0,1]$, $i=1,\dots,m$, of the $m\times m$ matrix $\rho_{A}$ are known as Schmidt numbers. 

The von Neumann entropy that describes the entanglement of the bipartite system is~\cite{Page93}
\begin{equation}\label{eq:vN}
S=-\tr\left(\rho_{A}\ln\rho_{A}\right)=-\sum_{i=1}^{m}\lambda_{i}\ln\lambda_{i}
\end{equation}
with the support $S\in[0,\ln m]$. Assuming independent standard complex Gaussian coefficients $c_{i,j}$, the joint eigenvalue density $f(\bm{\lambda})$ of the reduced density matrix $\rho_{A}$, also known as Hilbert-Schmidt ensemble~\cite{Page93,Majumdar}, is proportional to
\begin{equation}\label{eq:fte}
f(\bm{\lambda})\propto\delta\left(1-\sum_{i=1}^{m}\lambda_{i}\right)\prod_{1\leq i<j\leq m}\left(\lambda_{i}-\lambda_{j}\right)^{2}\prod_{i=1}^{m}\lambda_{i}^{\alpha},
\end{equation}
where $\delta(\cdot)$ is the Dirac delta function and 
\begin{equation}\label{eq:alpha}
\alpha=n-m
\end{equation}
denotes dimension difference\footnote{The densities~(\ref{eq:fte}) and~(\ref{eq:we}) are in fact valid for a non-negative real $\alpha$, and so are the subsequent results in this work. The parameter $\alpha$ will only be replaced by the dimension difference~(\ref{eq:alpha}) when displaying final cumulant expressions of $S$.} of the two subsystems.

Cumulants of entanglement entropy $S$ contain important statistical information of bipartite entanglement. In principle, the set of all cumulants uniquely determines the distribution of $S$ due to its compact support (Hausdorff's moment problem), where more accurate approximations to the distribution can be constructed with more higher-order cumulants obtained. In practice, the first two cumulants of mean and variance respectively specifies the typical value and fluctuation while higher-order cumulants govern tail behavior of the distribution. 

Due to the moment-cumulant relations~(\ref{eq:mk})--(\ref{eq:km}), one has the freedom to work with either set of moments or cumulants. As it becomes clear, it is of convenience to convert the moments (hence the cumulants) of entanglement entropy $S$ defined in~(\ref{eq:vN}) over the Hilbert-Schmidt ensemble~(\ref{eq:fte}) to the moments of an induced entropy
\begin{equation}\label{eq:von}
T=\sum_{i=1}^{m}x_{i}\ln x_{i},
\end{equation}
with $x_{i}\in[0,\infty)$ and the support $T\in[0,\infty)$, over Wishart-Laguerre ensemble~$g(\bm{x})$, whose joint density is proportional to~\cite{Mehta,Forrester,Page93}
\begin{equation}\label{eq:we}
g(\bm{x})\propto\prod_{1\leq i<j\leq m}\left(x_{i}-x_{j}\right)^{2}\prod_{i=1}^{m}x_i^{\alpha}\ee^{-x_i}.
\end{equation}
Moment conversion between the ensembles~(\ref{eq:fte}) and~(\ref{eq:we}) enables moment calculation since the Wishart-Laguerre ensemble~(\ref{eq:we}) is one of the most well-studied determinantal point processes~\cite{Mehta,Forrester}. Specific moment conversion formulas in computing the first four moments can be found respectively in~\cite{Page93,Wei17,Wei20,HWC21}. In the following, we provide a general conversion formula valid for moments of any order. 

\begin{lemma}\label{lemmast}
The $l$-th positive integer moment of $S$ can be recursively converted to the first $l$ moments of $T$ by 
\begin{equation}\label{eq:TSC}
\mathbb{E}\!\left[S^l\right]=(-1)^{l}\frac{\Gamma(mn)}{\Gamma(mn+l)}\mathbb{E}\!\left[T^l\right]+\sum_{j=0}^{l-1} A_{j}\mathbb{E}\!\left[S^j\right],
\end{equation}
where the coefficient $A_{j}$ is
\begin{equation}\label{eq:Aj}
A_{j}=(-1)^{j+l+1}\binom{l}{j}B_{l-j}\left(\psi_0(mn+l),\dots,\psi_{l-j-1}(mn+l)\right)
\end{equation}
with $\psi_{k}(z)$ and $B_k(z_1,\dots,z_k)$ respectively denoting the $k$-th polygamma functions~(\ref{eq:polygamma}) and the $k$-th complete exponential Bell polynomials~(\ref{eq:cbp2}).
\end{lemma}
The proof of Lemma~\ref{lemmast} is in Appendix~\ref{appst}. The task of calculating the moments of $S$ is now converted to calculating these of $T$ by repeated application of the conversion formula~(\ref{eq:TSC}) until only moments of $T$ remain. 

The starting point of cumulant computation for both existing and new methods is the following result derived in~\cite{Soshnikov}. The $l$-th cumulant $\kappa_l(X)$ of a linear statistics 
\begin{equation}\label{eq:X}
X=\sum_{i=1}^{m}f\hspace{-0.1em}\left(x_i\right)
\end{equation}
of a determinantal point process corresponding to a Hermitian random matrix ensemble is given by a sum of $l$ integrals as
\begin{equation}\label{eq:cuI}
\kappa_{l}(X)=\sum_{i=1}^{l}\hspace{0.1em}\mathrm{I}_{i},
\end{equation}
where the integrals are
\begin{equation}\label{eq:I}
\mathrm{I}_i=\sum_{l_1+\dots+l_i=l}\frac{(-1)^{i-1}}{i}\frac{l!}{l_1!\cdots l_i!}\int\prod_{j=1}^{i}f^{l_j}\!\left(x_j\right)K\!\left(x_j,x_{j+1}\right)\!\dd x_j
\end{equation}
with $K(x,y)$ being the correlation kernel of the ensemble and $x_{i+1}=x_{1}$. 

In our case, the linear statistics~(\ref{eq:X}) is $f(x)=x\ln x$ and the correlation kernel $K(x,y)$ of Wishart-Laguerre ensemble is given either by~(\ref{eq:ksum}) or~(\ref{eq:kernelcm}). The significance of~(\ref{eq:cuI}) is that it works out the underlying combinatorics to compactly represent the $l$-th cumulant into $l$ integrals, which otherwise involves, by the definition~(\ref{eq:Nei}), $l!$ number of integrals. 

\subsection{Existing results and methods}\label{sec1.2}
Computing exact cumulants of entanglement entropy $S$ over the Hilbert-Schmidt ensemble originated from the work of Page~\cite{Page93} in 1993, where he conjectured an expression of the first cumulant $\kappa_1(S)$. Page's conjecture was proved by S\'{a}nchez-Ruiz~\cite{Ruiz95} among several other proofs including~\cite{Foong94}. Some two decades later in 2016, a formula of the second cumulant $\kappa_2(S)$ was conjectured by Vivo, Pato, and Oshanin~\cite{VPO16}, which was proved in~\cite{Wei17}. Subsequently, formulas of higher-order cumulants $\kappa_3(S)$ and $\kappa_4(S)$ relevant to skewness and kurtosis were also obtained in~\cite{Wei20} and~\cite{HWC21}, respectively. We note that, besides the characterization via exact cumulants, large-dimensional behavior of entanglement entropy has been studied in~\cite{ASY13,HLW06}. 

Existing methods that led to the exact cumulants directly compute each of $l$ integrals~(\ref{eq:I}) in obtaining the $l$-th cumulant formula using a three-step process of decouple, compute, and simplify.
\begin{description}
\item \textbf{1. Decouple.} Since the $i$ variables $x_j$, $j=1,\dots,i$, of the integral $\mathrm{I}_i$ in~(\ref{eq:I}) are pairwise coupled through the kernel $K\!\left(x_j,x_{j+1}\right)$, the first step is to decouple $\mathrm{I}_i$ into a product of $i$ integrals of a single variable by replacing each kernel with its summation form~\cite{Szego,Forrester}
\begin{equation}\label{eq:ksum}
K\!\left(x,y\right)=\sqrt{w(x)w(y)}\sum_{k=0}^{m-1}\frac{k!}{(k+\alpha)!}L_{k}^{(\alpha)}(x)L_{k}^{(\alpha)}(y).
\end{equation}

\item {\bf 2. Compute.} The corresponding single integrals over the product of Laguerre polynomials $L_{k}^{(\alpha)}(x)$ and powers of linear statistics $f(x)=x\ln x$ are then evaluated by up to $l$-th derivative with respect to $q$ of the integral~\cite{Schr}
\begin{eqnarray}\label{eq:scho}
&&\int_{0}^{\infty}\!\!x^{q}\ee^{-x}L_{s}^{(a)}(x)L_{t}^{(b)}(x)\dd{x} \nonumber \\
&=&(-1)^{s+t}\sum_{k=0}^{\min(s,t)}{q-a\choose s-k}{q-b\choose t-k}\frac{\Gamma(q+1+k)}{k!},~~~~q>-1,
\end{eqnarray}
where the resulting integral identities contain the needed powers of $\ln x$ on the left-hand side and produce summations of polygamma functions~(\ref{eq:polygamma}) on the right-hand side.

\smallskip

\item {\bf 3. Simplify.} With all single integrals evaluated, each integral $\mathrm{I}_i$, $i=1,\dots,l$, now becomes an $i$-nested summation as a result of the decoupling procedure of the first step, where the last step is to simplify these summations involving polygamma functions. 
\end{description}

While the first two steps that convert integrals into summations can be performed straightforwardly, the bulk of calculation of existing methods lies in the third step of simplifying summations. The simplification of up to $l$-nested summations before arriving at cumulant expression $\kappa_{l}(T)$ is a case-by-case task, which becomes increasingly tedious as the order of cumulant $l$ increases. Specifically, new summation structures will arise as $l$ grows that require new insights for an appropriate change of summation order so as to evaluate $l$-nested sums one after one. At the same time, the number and types of summations also increase rapidly as $l$ grows, which constantly calls for new summation identities. In fact, the number of tailor-made summation identities needed increases from twelve for simplifying summations of the second cumulant~\cite{Wei17} to one hundred four\footnote{The arXiv version of the work~\cite{HWC21}, arXiv:2107.10978, contains a complete list of the one hundred four summation identities.} for that of the fourth cumulant~\cite{HWC21}. 

The difficulty of the simplification task is due in no small part to the emergence of unsimplifiable single summations of what we refer to as anomalies. The name anomaly reflects the somewhat unexpected fact that while final cumulant formulas admit closed-form expressions in the sense of no unsimplifiable summations involved, see for instance~(\ref{eq:k2}),~(\ref{eq:k3}),~(\ref{eq:k4}),~(\ref{eq:k5}),~(\ref{eq:k6}), their constitutes $\mathrm{I}_i$ do not. More precisely, as observed in~\cite{Wei17,Wei20,HWC21}, each individual nested summation $\mathrm{I}_i$ is simplified to closed-form terms plus a set of single summations of anomalies. These anomalies, however, cancel completely when summing up all $\mathrm{I}_i$ in~(\ref{eq:cuI}) leading to a closed-form cumulant formula in terms of polygamma functions. The number of anomalies also grows fast as the order of cumulant $l$ increases, where the corresponding identification and isolation of anomalies from nested summations become more complicated.

To illustrate the phenomenon of anomaly cancellation, we sketch the derivation~\cite{Wei17} of the second cumulant $\kappa_{2}(T)$ as an example. The general result~(\ref{eq:cuI}) states that $\kappa_{2}(T)$ is given by a sum of two integrals
\begin{equation}\label{eq:k2I}
\kappa_{2}(T)=\mathrm{I}_{1}+\mathrm{I}_{2},
\end{equation}
where the integrals are~(\ref{eq:I})
\begin{eqnarray}
\mathrm{I}_1 &=& \int_{0}^{\infty}x_{1}^2\ln^{2}x_{1}~\!K(x_1,x_1)\dd x_1, \\
\mathrm{I}_2 &=& -\int_{0}^{\infty}\!\!\!\int_{0}^{\infty}x_{1}x_{2}\ln x_1\ln x_2~\!K(x_1,x_2)K(x_2,x_1)\dd x_1 \dd x_2.
\end{eqnarray}
Applying the three-step process of existing methods of decoupling the integrals via~(\ref{eq:ksum}), computing the decoupled single integrals using~(\ref{eq:scho}) and its first two derivatives, and simplifying laboriously the resulting single sums in~$\mathrm{I}_1$ and double sums in~$\mathrm{I}_2$ by the twelve summation identities in~\cite{Wei17} leads eventually to
\begin{eqnarray}
\mathrm{I}_{1} &=& a\Omega_1+\Psi_1, \label{eq:I1} \\
\mathrm{I}_{2} &=& -a\Omega_1+\Psi_2, \label{eq:I2}
\end{eqnarray}
where $\Psi_1$, $\Psi_2$ denote closed-form terms omitted and
\begin{equation}\label{eq:aT2}
\Omega_1=\sum_{k=1}^{m}\frac{\psi_{0}(k+\alpha)}{k}    
\end{equation}
is the only anomaly in calculating $\kappa_{2}(T)$. This anomaly naturally appears in simplifying $\mathrm{I}_1$, whereas its identification and isolation from $\mathrm{I}_2$ double summations are fairly demanding. Nevertheless, the anomaly~(\ref{eq:aT2}) cancels when summing up $\mathrm{I}_{1}$ and $\mathrm{I}_{2}$ in~(\ref{eq:k2I}) leading to the closed-form formula of $\kappa_{2}(T)$ as shown in~(\ref{eq:T2}). 

The number of anomalies increases rapidly to seventeen in $\kappa_{4}(T)$ computation, a partial list of which is found in Table~\ref{table1}. The seventeen anomalies are disguised in up to quadruple summations stemmed from computing the integrals~(\ref{eq:I}), where identifying the anomalies out of the summations becomes a non-trivial task. The subsequent task of isolating these anomalies by evaluating the nested summations via one hundred four tailor-made summation identities is highly laborious. In fact, it took some two years for the authors~\cite{HWC21} to complete the simplification task, where it does not seem viable to derive cumulant formulas beyond $\kappa_{4}(T)$ using existing methods. 

\begin{table}[h!]
\vspace{0.1cm}
\centering\ra{1}
{\renewcommand{\arraystretch}{2.4}
\begin{tabular}{||l|l|l||}
\Xhline{0.8pt}
\scalebox{0.9}{$\displaystyle\small\Omega_1=\sum_{ k=1}^m \frac{\psi_0(k+\alpha)}{k}$} & \scalebox{0.9}{$\displaystyle\Omega_{6}=\sum_{k=1}^m \frac{\psi_0(k)\psi_0(k+\alpha)}{k}$} & \scalebox{0.9}{$\displaystyle\Omega_{11}=\sum_{k=1}^m \frac{\psi_1(k+\alpha)}{k+\alpha}$} \\ 
\scalebox{0.9}{$\displaystyle\Omega_2=\sum_{k=1}^m \frac{\psi_0(k+\alpha)}{k^2}$} & \scalebox{0.9}{$\displaystyle \Omega_7=\sum_{k=1}^m \frac{\psi_0^3(k+\alpha)}{k}$} & \scalebox{0.9}{$\displaystyle \Omega_{12}=\sum_{k=1}^m \frac{\psi_0(k)\psi_1(k+\alpha)}{k}$} \\
\scalebox{0.9}{$\displaystyle\Omega_3=\sum_{k=1}^m\frac{\psi_0^2(k+\alpha)}{k}$} & \scalebox{0.9}{$\displaystyle \Omega_8=\sum_{k=1}^m \frac{\psi_0^3(k+\alpha)}{k+\alpha}$} & \scalebox{0.9}{$\displaystyle \Omega_{13}=\sum_{k=1}^m \frac{\psi_0(k+\alpha)\psi_1(k+\alpha)}{k}$} \\
\scalebox{0.9}{$\displaystyle\Omega_{4}=\sum_{k=1}^m\frac{\psi_0^2(k+\alpha)}{k+\alpha}$} & \scalebox{0.9}{$\displaystyle \Omega_{9}=\sum_{k=1}^m \frac{\psi_0(k)\psi_0^2(k+\alpha)}{k}$} & \scalebox{0.9}{$\displaystyle \Omega_{14}=\sum_{k=1}^m \frac{\psi_2(k+\alpha)}{k}$} \\
\scalebox{0.9}{$\displaystyle\Omega_{5}=\sum_{k=1}^{m}\frac{\psi_0^2(k+\alpha)}{k^2}$} &\scalebox{0.9}{$\!\!\displaystyle \Omega_{10}=\sum_{k=1}^m \frac{\psi_1(k+\alpha)}{k}$} &\scalebox{0.9}{$\displaystyle\Omega_{15}=\sum_{k=1}^{m}\frac{\psi_2(k+\alpha)}{k+\alpha}$} \\ [2 mm]
\Xhline{0.8pt}
\end{tabular}}
\vspace{0.4cm}
\caption{List of anomalies in the simplification of $\kappa_4(T)$.} \label{table1}
\end{table}

The complication of existing methods originates from the decoupling procedure with each use of~(\ref{eq:ksum}) introducing one more layer of nested summations, from which anomalies emerge. The subsequent complete cancellation indicates that anomalies and, therefore, the decoupling procedure may not be indispensable in the cumulant computation. A natural question is whether a new decoupling procedure can be found to avoid simplifying summations so as also to obstruct the onset of anomalies. As presented in the next section, the key finding of this work is a summation-free decoupling procedure based on the discovered cumulant structures. The new cumulant computation methods guarantee the existence of a closed-form formula of cumulant of any order, which confirms that the appearance of anomalies is an artifact of existing summation-based methods. Importantly, the new methods generate cumulant formulas of any order in a straightforward manner while circumventing entirely the need to simplify summations. In particular, instead of two years~\cite{HWC21}, it now takes a few seconds to produce the cumulant formula of $\kappa_{4}(T)$ when implementing the new decoupling procedure in a computer algebra system.

The existing summation-based methods have also been applied to study entanglement entropy over Bures-Hall ensemble~\cite{Bortola14,FK16,SK2019,Wei20BHA} and fermionic Gaussian ensemble~\cite{BHK21,BHKRV22}. For these two ensembles, the simplification of summations leading to the respective variance formulas~\cite{Wei20BH,HW23} was no less tedious than that of Hilbert-Schmidt ensemble with the anomaly cancellation phenomenon persists. Finding cumulant structures of the two ensembles would be more desirable to compute their higher-order cumulants beyond variances.

The rest of the paper is organized as follows. In Section~\ref{sec2}, we outline new cumulant computation methods before presenting detailed cumulant structural results. Proofs of the results are found in Appendices. Examples of cumulant formulas up to the sixth cumulant computed using new methods are provided in Section~\ref{sec3}. 
  
\section{Cumulant Structures}\label{sec2}
In this section, we present new methods of cumulant computation of entanglement entropy based on the derived cumulant structures. Main ideas of new methods are discussed first in Section~\ref{sec2.1}. We then provide underlying matrix-level and kernel-level structural results in Section~\ref{sec2.2} before stating the main result on cumulant decoupling structure in Section~\ref{sec2.3}. 

\subsection{Overview of new methods}\label{sec2.1}
The new methods, summarized in Theorem~\ref{Theorem1}, provide a summation-free decoupling procedure that uncovers the structure of $l$-th cumulant $\kappa_l(T)$ as a function of joint cumulants involving two families of ancillary linear statistics
\begin{eqnarray}
T_k&=&\sum_{i=1}^{m}x_{i}^{k}\ln x_{i}, \label{eq:tk} \\
R_k&=&\sum_{i=1}^{m}x_{i}^{k}. \label{eq:rk}
\end{eqnarray}
The proposed decoupling procedure consists of a new decoupling structure enabled by matrix-level\footnote{Matrix-level and kernel-level results refer to cumulant structures derived with the knowledge of the ensemble's matrix-variate density~(\ref{eq:bwem}) and eigenvalue density~(\ref{eq:Nei}) in terms of correlation kernel, respectively.} results and a corresponding decoupled term recycled into lower-order cumulants by kernel-level results. Specifically, 
\begin{itemize}
\item The matrix-level results, Proposition~\ref{propdab}, enable the construction of a closely related but simpler joint cumulant referred to as decoupling statistics~(\ref{eq:destat}) that leads to a new decoupling structure~(\ref{eq:th1}) through the use of Christoffel-Darboux kernel~\cite{Szego,Forrester} 
\begin{eqnarray}\label{eq:kernelcm}
\!\!\!\!K\!\left(x,y\right)=\frac{m!\sqrt{w(x)w(y)}}{(m-1+\alpha)!}~\frac{L_{m-1}^{(\alpha)}(x)L_m^{(\alpha)}(y)-L_m^{(\alpha)}(x)L_{m-1}^{(\alpha)}(y)}{x-y}.
\end{eqnarray} 

\item The kernel-level results are more delicate tools to recycle the decoupled term~(\ref{eq:dT}) produced from the new decoupling structure~(\ref{eq:th1}) into lower-order cumulants. The decoupled term consists of three types of integrals involving kernels: the recycling of the first two types is presented in Proposition~\ref{props} and that of the third type is found in Proposition~\ref{propdk} and Corollary~\ref{codkb}.
\end{itemize}

At the heart of new methods is the discovery of the deep but subtle recursive nature of cumulants $\kappa_{l}(T)$. Unlike existing methods that seek to explicitly compute each constituent integral~(\ref{eq:I}) of $\kappa_{l}(T)$, the new methods aim to relate cumulants as a whole to lower-order ones. A key enabler is a summation-free decoupling through the kernel~(\ref{eq:kernelcm}) that fundamentally prevents from arising the illusive tasks intrinsic to existing methods of simplifying summations and canceling anomalies. Equally important as the new decoupling structure is the set of tailored tools to reconstruct cumulants from decoupled integrals. Iterations between decoupling and reconstruction drive the cumulant computation of new methods.

The new methods require the initial data of mean formulas of ancillary linear statistics~(\ref{eq:tk}) and~(\ref{eq:rk}) recursively obtained from the results below.
\begin{lemma}\label{lemmart}
Using the notation~(\ref{eq:lm3mm}), the recurrence relations of mean formulas $\kappa\!\left(R_k\right)$ and $\kappa\!\left(T_k\right)$ valid for $k\in\mathbb{R}_{\ge0}$ are respectively
\begin{equation}\label{eq:rrecur}
(k+1)\kappa\!\left(R_k\right)=(k-1)(2m+\alpha)\kappa\!\left(R_{k-1}\right)+m(m+\alpha)\left(\kappa^{\normalfont\plus}\!\left(R_{k-1}\right)-\kappa^{\normalfont\minus}\!\left(R_{k-1}\right)\right)
\end{equation}
and
\begin{eqnarray}\label{eq:Tkrecur}
(k+1)\kappa\!\left(T_{k}\right)&=&(k-1)(2m+\alpha)\kappa\!\left(T_{k-1}\right)+m(m+\alpha)\left(\kappa^{\normalfont\plus}\!\left(T_{k-1}\right)- \kappa^{\normalfont\minus}\!\left(T_{k-1}\right)\right) \nonumber \\
&&-~\!\kappa\!\left(R_{k}\right)+(2m+\alpha)\kappa\!\left(R_{k-1}\right),
\end{eqnarray}
where the initial values are
\begin{eqnarray}
\kappa(R_0) &=& m, \label{eq:iR} \\ 
\kappa(T_0) &=& (m+\alpha)\psi_0(m+\alpha)-\alpha\psi_{0}(\alpha)-m. \label{eq:iT}
\end{eqnarray}
\end{lemma}
The proof of Lemma~\ref{lemmart} is in Appendix~\ref{apprt}. For any positive integer $k$, explicit mean expressions $\kappa(R_k)$ and $\kappa(T_k)$ can now be derived, although the results~(\ref{eq:rrecur}) and~(\ref{eq:Tkrecur}) are valid for nonnegative real $k$. For $k=1$, one immediately computes from~(\ref{eq:Tkrecur}) the mean value of $T$ as
\begin{equation}\label{eq:T1}
\kappa(T)=m(m+\alpha)\psi_{0}(m+\alpha+1)+\frac{1}{2}m(m-1),
\end{equation}
which after converting to that of $S$ via Lemma~\ref{lemmast} recovers the celebrated Page's formula~\cite{Page93,Ruiz95,Foong94} of average von Neumann entropy
\begin{equation}\label{eq:S1}
\kappa(S)=\psi_{0}(mn+1)-\psi_{0}(n+1)-\frac{m-1}{2n}.
\end{equation}
For $k\ge2$, expressions of $\kappa(T_k)$ serve as starting point of new cumulant computation methods. As demonstrated in Section~\ref{sec3}, the mean formulas from $\kappa(T_2)$ to $\kappa(T_6)$ shown in~(\ref{eq:mT2}),~(\ref{eq:mT3}),~(\ref{eq:mT4}),~(\ref{eq:mT5}),~(\ref{eq:mT6}) are utilized as the first step of decoupling procedure in obtaining the respective cumulant expressions.

\subsection{Matrix-level and kernel-level structures}\label{sec2.2}
The following definitions are presented first. We denote ${\bf X}=\{X_1,\dots,X_l\}$ as a set of $l$ linear statistics
\begin{equation}
X_j=\sum_{i=1}^{m}f_{j}\hspace{-0.1em}\left(x_{i}\right).
\end{equation}
The corresponding joint cumulant of ${\bf X}$ is denoted by
\begin{equation}\label{eq:kld}
\kappa_l\!\left({\bf X}\right)
\end{equation}
whenever its generating function~(\ref{eq:ktg}), that uniquely determines the joint cumulant~(\ref{eq:kld}) through~(\ref{eq:ktd}), exists. The subscript $l$ is omitted if the number of variables in ${\bf X}$ is clear. For $l$ identical linear statistics $X=X_{1}=\dots=X_{l}$, the definition~(\ref{eq:kld}) reduces to that of the $l$-th cumulant of $X$. We also define
\begin{equation}\label{eq:lm3mm}
\kappa^{\substack{\normalfont\plus\vspace{-0.24em} \\ \normalfont\minus}}_l\!\left({\bf X}\right)=\kappa_l\!\left({\bf X}\right)\rvert_{m\to m\pm1}.
\end{equation}

To discuss matrix-level results below, it is necessary to work with a generalized Wishart density~(\ref{eq:bwem}) with parameters $\alpha,\beta\in\mathbb{R}_{\ge0}$, whose joint eigenvalue density reduces to the considered Wishart-Laguerre ensemble~(\ref{eq:we}) when $\beta=1$ and $\alpha=n-m$.
\begin{proposition}\label{propdab}
For linear statistics ${\bf X}$ of the Wishart density~(\ref{eq:bwem}), the joint cumulant $\kappa_l\!\left({\bf X}\right)$ satisfies the derivative relation
\begin{equation}\label{eq:matda}
\frac{\dd}{\dd\alpha}\kappa_l\!\left({\bf X}\right)=\kappa_{l+1}\!\left({\bf X},T_{0}\right).
\end{equation}
\end{proposition}
The proof of Proposition~\ref{propdab} is in Appendix~\ref{appa:propdab}. Another useful derivative relation is
\begin{equation}\label{eq:matdb}
\frac{\dd}{\dd\beta}\kappa_l\!\left({\bf X}\right)=-\kappa_{l+1}\!\left({\bf X},R\right),
\end{equation}
the proof of which is outlined also in Appendix~\ref{appa:propdab}. Proposition~\ref{propdab} permits the possibility to construct the decoupling statistics~(\ref{eq:destat}) such that the difference between the desired cumulant and the constructed one, i.e. the decoupling structure~(\ref{eq:th1}), decouples the kernels through the Christoffel-Darboux form~(\ref{eq:kernelcm}). 

Before discussing kernel-level results, we need the following combinatorial structure of joint cumulant $\kappa_l\!\left({\bf X}\right)$ in terms of kernels.
\begin{lemma}\label{lemmac}
The joint cumulant $\kappa_l\!\left({\bf X}\right)$ can be represented as
\begin{equation}
\kappa_l\!\left({\bf X}\right)=\sum_{\left\{p_1,\dots, p_i\right\}\in\mathcal{P}_{L}}\sum_{\sigma\in S_i}\frac{(-1)^{i-1}}{i}\int_{[0,\infty)^i}\prod_{j=1}^{i}F_jK\!\left(x_j,x_{j+1}\right)\dd x_j, \label{eq:jcuker}
\end{equation}
where $\mathcal{P}_{L}$ is the set of partitions of $L=\{1,\dots,l\}$ with $\bigcup_{j=1}^{i} p_j=L$, $\sigma$ is an element of symmetric group $S_i$, and $F_j=\prod_{r\in p_j}f_r\!\left(x_{\sigma\left(j\right)}\right)$. 
\end{lemma}

While Lemma~\ref{lemmac} generalizes the result~(\ref{eq:cuI}) in~\cite{Soshnikov} of joint cumulants of $l$ identical linear statistics to $l$ arbitrary linear statistics, the idea of its proof is rather similar to that of~(\ref{eq:cuI}). We nevertheless include the proof of Lemma~\ref{lemmac} in Appendix~\ref{appa:lemmac} for completeness. 

We now introduce kernel-level results that recast integrals resulting from the new decoupling structure~(\ref{eq:th1}) into lower-order joint cumulants of ancillary statistics~(\ref{eq:tk}) and~(\ref{eq:rk}). Integrals that give rise to the decoupled term~(\ref{eq:dT}) consist of three types $H_l({\bf X})$, $h_l({\bf X})$, and $D_l({\bf X})$ as defined in~(\ref{eq:dcIth}) in the proof of Theorem~\ref{Theorem1}, which are also summarized in Table~\ref{table2}. 

\begin{table}[h]
\centering\ra{1.8}
\begin{tabular}{c c}
\noalign{\hrule height 0.7pt}
Decoupled integrals~(\ref{eq:dcIth}) & $L(x,y)$ \\
\hline 
\eqmakebox[log][r]{$H_l\!\left({\bf{X}}\right)$} & $L_{\hspace{-0.1em}H}(x, y)$ in~(\ref{eq:LH}) \\ 
\eqmakebox[log][r]{$h_l\!\left({\bf{X}}\right)$} & $L_h(x, y)$ in~(\ref{eq:Lh}) \\
\eqmakebox[log][r]{$D_l\!\left({\bf{X}}\right)$} & $L_{\hspace{-0.1em}D}(x,y)$ in~(\ref{eq:LD}) \\
\noalign{\hrule height 0.7pt}
\end{tabular}
\vspace{0.4cm}
\caption{Three types of decoupled integrals.}\label{table2}
\end{table}

Kernel-level results that recycle the integrals $H_l\!\left({\bf{X}}\right)$ and $h_l\!\left({\bf{X}}\right)$ are presented first. 
\begin{proposition}\label{props}
The integrals $H_l\!\left({\bf{X}}\right)$ and $h_l\!\left({\bf{X}}\right)$ are recast respectively to lower-order cumulants as
\begin{eqnarray}
H_l({\bf X})&=&\sum_{\left\{p_1,\dots, p_i\right\}\in\mathcal{P}_{L}}\prod_{j=1}^{i}\left(\kappa_{\rvert p_j\rvert}^{\normalfont\plus}\!\left({\bf X}_{p_j}\right)-\kappa_{\rvert p_j\rvert}\!\left({\bf X}_{p_j}\right)\right), \label{eq:Hcs} \\
h_l({\bf X})&=&-\sum_{\left\{p_1,\dots, p_i\right\}\in\mathcal{P}_{L}}\prod_{j=1}^{i}\left(\kappa_{\rvert p_j\rvert}^{\normalfont\minus}\!\left({\bf X}_{p_j}\right)-\kappa_{\rvert p_j\rvert}\!\left({\bf X}_{p_j}\right)\right), \label{eq:hcs}
\end{eqnarray}
where ${\bf X}_{p_j}=\{X_r: r\in p_j\}$ with $\bigcup_{j=1}^i{\bf X}_{p_j}={\bf X}$.
\end{proposition}

The proof of Proposition~\ref{props} is in Appendix~\ref{appa:props}. Recycling of the decoupled integral $D_l\!\left({\bf{X}}\right)$ is represented via joint cumulant derivative defined as
\begin{equation}\label{eq:notationdk}
\kappa_l^{\prime}({\bf X}) =\kappa\big(X_1^\prime,\dots,X_l\big)+\kappa\big(X_1,X_2^\prime,\dots,X_l\big)+\dots+\kappa\big(X_1,\dots,X_l^\prime\big),
\end{equation}
where
\begin{equation}\label{eq:xprime}
X_{j}^{\prime}=\sum_{i=1}^{m}x_{i}\frac{\dd}{\dd x_i}f_{j}(x_i).
\end{equation}    

\begin{proposition}\label{propdk}
The integral $D_l\!\left({\bf{X}}\right)$ is recast to lower-order cumulants as
\begin{equation}\label{eq:redk} 
D_l({\bf X})=\kappa^\prime_l({\bf X}). \\
\end{equation}
\end{proposition}

The proof of Proposition~\ref{propdk} is found in Appendix~\ref{appa:propdk}. As a consequence of Proposition~\ref{propdk}, the integral $D_l({\bf X})$ reduces further to summations of lower-order cumulants in two special cases of joint statistics ${\bf X}=\{T_k,T,\dots,T\}$ and ${\bf X} = \{R_k,T,\dots,T\}$ useful in the subsequent calculation as summarized below.

\begin{corollary}\label{codkb}
\begin{eqnarray}
D_l\!\left(T_k,T,\dots,T\right)
&=&\sum_{i=0}^{l-1}\frac{(l-1)!}{(l-1-i)!}\big(\hspace{0.1em}(k+l-1-i)\kappa_{l-i}\!\left(T_k,T,\dots,T\right) \nonumber \\
&&+~\!\kappa_{l-i}\!\left(R_k,T,\dots,T\right)\hspace{-0.1em}\big), \label{eq:coro11} \\
D_l\!\left(R_k,T,\dots,T\right)&=&\sum_{i=0}^{l-1}\frac{(l-1)!}{(l-1-i)!}(k+l-1-i)\kappa_{l-i}\!\left(R_k,T,\dots,T\right).\label{eq:coro12}
\end{eqnarray}
\end{corollary}
The proof of Corollary~\ref{codkb} is in Appendix~\ref{appa:coro1}. For later use, we introduce shorthand notations $H_{l,s}(k)$ and $D_{l,s}(k)$,
\begin{eqnarray}
H_{l,s}(k)&=&\sum_{r=0}^{k-1}H_s(T_r,T,\dots,T)h_{l-s}(T_{k-r-1},T,\dots,T), \label{eq:dcs1} \\
D_{l,s}(k)&=&\sum_{r=0}^{k-1}D_s(T_r,T,\dots,T)D_{l-s}(T_{k-r-1},T,\dots,T), \label{eq:dcs2}
\end{eqnarray}
which, according to Proposition~\ref{props} and Corollary~\ref{codkb}, comprise joint cumulants of order no more than $l-1$. Therefore, the terms $H_{l,s}(k)$ and $D_{l,s}(k)$ are explicitly available once the corresponding lower-order cumulants have been obtained. 

\subsection{Main results}\label{sec2.3}
A proper integration of matrix-level and kernel-level results leads to the following main result of this work on a summation-free cumulant decoupling procedure of entanglement entropy. 

\begin{theorem}\label{Theorem1} 
For any $l\geq2$, the joint cumulant $\kappa_l(T_k,T,\dots,T)$ admits the decoupling structure
\begin{equation}\label{eq:th1}
\kappa_l(T_k,T,\dots,T)-\frac{\dd}{\dd\alpha}\kappa_{l-1}(T_{k+1},T,\dots,T)=\delta_l(k),
\end{equation}
where the decoupled term
\begin{equation}\label{eq:dT}
\delta_l(k)=\sum_{s=1}^{l-1}\frac{(l-2)!}{(s-1)!(l-s-1)!}\left(\kappa(R)H_{l,s}(k)-D_{l,s}(k)\right)
\end{equation} 
consists of lower-order cumulants $H_{l,s}(k)$ and $D_{l,s}(k)$ defined in~(\ref{eq:dcs1}) and~(\ref{eq:dcs2}), respectively.
\end{theorem}

The proof of Theorem~\ref{Theorem1} is in Appendix~\ref{appa:Theorem1}. A few remarks are in order. \\

\noindent{\bf\small Remarks}
\begin{itemize}
\item In addition to the decoupling structure of $T_k$ in~(\ref{eq:th1}), the proof of Theorem~\ref{Theorem1} also requires the decoupling structure of joint cumulants $\kappa_l(R_k,T,\dots,T)$ involving $R_k$, which can be similarly determined as shown in~(\ref{eq:th2d}).

\item Theorem~\ref{Theorem1} implies the existence of a closed-form cumulant formula $\kappa_l(T)$ for any order $l$. This is seen inductively in $l$ from the decoupling structure~(\ref{eq:th1}) between consecutive cumulants, where lower-order cumulants in the decoupled term~(\ref{eq:dT}) are closed-form expressions 
by the induction hypothesis. The base case that the mean value $\kappa(T_l)$ admits a closed-form expression is verified by Lemma~\ref{lemmart}. The existence in turn implies the presence of anomalies and the use of polygamma summation identities essential in existing methods are not necessary. 

\item The existence guaranteed by Theorem~\ref{Theorem1} also provides an explicit construction in generating the closed-form expression of $\kappa_l(T)$ for a given $l$. As illustrated in the pseudocode\footnote{Algorithm~\ref{alg} has been implemented in the computer algebra system~\textsc{Mathematica}. The codes are available upon reasonable requests to any of the authors.} Algorithm~\ref{alg} below, computing $\kappa_l(T)$ as the output requires the input of cumulant order $l$ and the input of mean formula $\kappa(T_l)$ from Lemma~\ref{lemmart}. The calculation from $\kappa\!\left(T_l\right)$ to $\kappa_l(T)$ comprises $l-1$ iterations as shown in the while loop from line $2$ to line $7$. In each iteration, the decoupled term~(\ref{eq:dT}) is calculated first in line $4$ before the decoupling structure~(\ref{eq:th1}) is employed to obtain the next order cumulant in line $5$.

\begin{algorithm}
\caption{Calculating $l$-th Cumulant $\kappa_l(T)$}\label{alg}
\hspace*{\algorithmicindent}\textbf{Input:} Any positive integer $l\geq2$ \\ 
\hspace*{\algorithmicindent}\hspace{3.11em} $\kappa(T_l)$ expression of Lemma~\ref{lemmart} \\ 
\hspace*{\algorithmicindent}\textbf{Output:} Closed-form formula of $\kappa_l(T)$ 
\begin{algorithmic}[1]
\State $L \gets 2$
\While{$L\leq l$}
\State $k \gets l-L+1$
\State $\delta_L(k) \gets$ (\ref{eq:dT}) of Theorem~\ref{Theorem1}
\State $\kappa_L\!\left(T_{k},T,\dots,T\right) \gets \delta_L(k)+\frac{\dd}{\dd\alpha}\kappa_{L-1}\!\left(T_{k+1},T,\dots,T\right)$ 
\State $L \gets L+1$
\EndWhile
\end{algorithmic}
\end{algorithm}
\end{itemize}

As a consequence of Theorem~\ref{Theorem1} and Lemma~\ref{lemmast}, we have the following result.
\begin{corollary}\label{coro2}
In the expression of $l$-th cumulant $\kappa_l(S)$ of von Neumann entropy, the terms involving polygamma function of highest order $\psi_{l-1}$ are
\begin{equation}\label{eq:coro2}
(-1)^{l-1}\left(\psi_{l-1}(mn)-\frac{\kappa\!\left(R_l\right)}{(mn)_{l}}\psi_{l-1}(n)\right).
\end{equation}
\end{corollary}

The proof of Corollary~\ref{coro2} is in Appendix~\ref{appa:coro2}. Despite that Algorithm~\ref{alg} is able to generate the full expression of $\kappa_l(S)$ for a given $l$ with the help of Lemma~\ref{lemmast}, it is unable to provide the analytical expression of highest-order polygamma term valid for any $l$ as captured in Corollary~\ref{coro2}. 

\section{Cumulant Calculation via New Methods}\label{sec3}
In this section, we present examples of cumulant computation of entanglement entropy utilizing new methods of cumulant structures. In Section~\ref{sec:3.1}, we rederive cumulant formulas up to the fourth order\footnote{The calculation of mean formula~(\ref{eq:S1}), that does not require the decoupling procedure of Theorem~\ref{Theorem1}, is excluded here.} including variance, skewness, and kurtosis, these formulas were obtained in the literature through tedious existing methods. Examples of further higher-order cumulant expressions, that are out of reach in practice by existing methods, are provided in Section~\ref{sec:3.2}. 

\subsection{Reproducing existing results: $\kappa_2$, $\kappa_3$, and $\kappa_4$}\label{sec:3.1}
\subsubsection*{Calculation of $\kappa_2$}
Following the implementation of Theorem~\ref{Theorem1} via Algorithm~\ref{alg}, the calculation of second cumulant $\kappa_2(T)$ consists of one iteration of computation between decoupled term~(\ref{eq:dT}) and decoupling structure~(\ref{eq:th1}) as
\begin{equation}\label{eq:k2i1}
\kappa_2\!\left(T\right)=\delta_2(1)+\frac{\dd}{\dd\alpha}\kappa(T_{2}).
\end{equation}
The iteration~(\ref{eq:k2i1}) requires the mean
\begin{equation}\label{eq:mT2}
\kappa(T_2)=\kappa(R_2)\psi_0(m+\alpha)+\Lambda_2
\end{equation}
with
\begin{eqnarray}
\kappa(R_2) &=& m(m+\alpha)(2m+\alpha), \label{eq:kR2} \\
\Lambda_2 &=& \frac{m}{6}\left(10m^{2}+9m\alpha+6m+3\alpha+2\right)
\end{eqnarray}
obtained from Lemma~\ref{lemmart}, cf. the structure~(\ref{eq:tl}), and the decoupled term 
\begin{equation}\label{eq:meanth1}
\delta_2(1)=\kappa(R)H_{2,1}(1)-D_{2,1}(1)
\end{equation}
with lower-order cumulants 
\begin{equation}\label{ex:H1}
H_{2,1}(1)=\left(\kappa^\plus\!\left(T_0\right)-\kappa\!\left(T_0\right)\right)\left(\kappa\!\left(T_{0}\right)-\kappa^\minus\!\left(T_{0}\right)\right)  
\end{equation}
and
\begin{equation}\label{ex:D1}
D_{2,1}(1)=\kappa^2\!\left(R_0\right)    
\end{equation}
computed by Proposition~\ref{props} and Proposition~\ref{propdk} along with Corollary~\ref{codkb}, respectively, when using the definitions~(\ref{eq:dcs1}) and~(\ref{eq:dcs2}). Putting the results together, we arrive at the desired cumulant structure
\begin{equation}\label{ex:dsk2}
\kappa_2\!\left(T\right)=\kappa(R)\left(\kappa^\plus\!\left(T_0\right)-\kappa\!\left(T_0\right)\right)\left(\kappa\!\left(T_{0}\right)-\kappa^\minus\!\left(T_{0}\right)\right)-\kappa^2\!\left(R_0\right)+\frac{\dd}{\dd\alpha}\kappa(T_{2}).
\end{equation}

Inserting lower-order cumulant expressions available hitherto in the cumulant structure~(\ref{ex:dsk2}) leads to an explicit formula of $\kappa_2\!\left(T\right)$ as
\begin{eqnarray}\label{eq:T2}
\kappa_2\!\left(T\right)&=&m(m+\alpha)(2m+\alpha)\psi_1(m+\alpha+1)+m(m+\alpha)\psi_{0}^{2}(m+\alpha+1) \nonumber \\
&&+~\!m(3m+2\alpha-1)\psi_{0}(m+\alpha+1)+\frac{1}{2}m(m-1).
\end{eqnarray}
Converting the resulting formula to that of $S$ via Lemma~\ref{lemmast} while keeping in mind the relation~(\ref{eq:alpha}), we obtain the second cumulant formula of entanglement entropy
\begin{equation}\label{eq:k2}
\kappa_2(S)=-\psi_1(mn+1)+\frac{m+n}{mn+1}\psi_1(n+1)-\frac{(m-1)(m+2n-1)}{4n^2(mn+1)}.
\end{equation}
It is worth mentioning that for $l=2$ Corollary~\ref{coro2} gives
\begin{equation}
-\psi_{1}(mn)+\frac{\kappa\!\left(R_2\right)}{(mn)_{2}}\psi_{1}(n),
\end{equation}
which after applying~(\ref{eq:kR2}) and~(\ref{eq:psisre}) recovers terms involving highest-order polygamma function $\psi_{1}$ in~(\ref{eq:k2}) as expected.

The cumulant formula $\kappa_2(S)$ in~(\ref{eq:k2}) was first derived in~\cite{Wei17} using the existing summation-based methods. Compared to the twelve-page proof in~\cite{Wei17}, the new methods based on cumulant structures lead to the above-presented proof of about one page. In particular, tasks in~\cite{Wei17} of simplifying summations using a dozen of tailor-made summation identities while identifying and isolating one anomaly, $\Omega_1$ in Table~\ref{table1}, from the summations are completely avoided. 

\subsubsection*{Calculation of $\kappa_3$}
Following the implementation of Theorem~\ref{Theorem1} via Algorithm~\ref{alg}, the calculation of third cumulant $\kappa_3(T)$ consists of two iterations of computation between decoupled term~(\ref{eq:dT}) and decoupling structure~(\ref{eq:th1}) as
\begin{eqnarray}
\kappa\!\left(T_2,T\right) &=& \delta_2(2)+\frac{\dd}{\dd\alpha}\kappa(T_3), \label{eq:k3i1} \\
\kappa_3(T) &=& \delta_3(1)+\frac{\dd}{\dd\alpha}\kappa(T_2,T). \label{eq:k3i2}
\end{eqnarray}
The first iteration~(\ref{eq:k3i1}) requires the mean 
\begin{equation}\label{eq:mT3}
\kappa(T_3)=\kappa(R_3)\psi_0(m+\alpha)+\Lambda_3
\end{equation}
with
\begin{eqnarray}
\kappa(R_3)&=&m(m+\alpha)\left(5m^2+5m\alpha+\alpha^2+1\right), \label{eq:kR3} \\
\Lambda_3&=&\frac{m}{12}\big(59m^3+88m^2\alpha+30m^2+30m\alpha^2+30m\alpha+37m \nonumber \\
&&+~\!6\alpha^2+26\alpha+6\big)
\end{eqnarray}
obtained from Lemma~\ref{lemmart}, cf. the structure~(\ref{eq:tl}), where $\kappa\!\left(T_2,T\right)$ becomes available after computing the decoupled term $\delta_2(2)$ by~(\ref{eq:dT}). Taking $\kappa\!\left(T_2,T\right)$ as input, the second iteration~(\ref{eq:k3i2}) leads to the desired cumulant structure of $\kappa_3(T)$ by computing the decoupled term $\delta_3(1)$ using~(\ref{eq:dT}) as 
\begin{eqnarray}\label{ex:dt3}
\delta_3(1)&=&\kappa(R)H_{3,1}(1)-D_{3,1}(1)+\kappa(R)H_{3,2}(1)-D_{3,2}(1)
\end{eqnarray}
with lower-order cumulants 
\begin{eqnarray}
H_{3,1}(1)&=&\big(\kappa\!\left(T,T_0\right)-\kappa^{\minus}\!\left(T,T_0\right)-\left(\kappa\!\left(T_0\right)-\kappa^{\minus}\!\left(T_0\right)\right)\left(\kappa\!\left(T\right)-\kappa^{\minus}\!\left(T\right)\right)\!\big) \nonumber \\
&&\times\left(\kappa^{\plus}\!\left(T_0\right)-\kappa\!\left(T_0\right)\right), \\
H_{3,2}(1)&=&\big(\kappa^{\plus}\!\left(T,T_{0}\right)-\kappa\!\left(T,T_{0}\right)+\left(\kappa^{\plus}\!\left(T_{0}\right)-\kappa\!\left(T_{0}\right)\right)\left(\kappa^{\plus}\!\left(T\right)-\kappa\!\left(T\right)\right)\!\big) \nonumber \\
&&\times\left(\kappa\!\left(T_{0}\right)-\kappa^{\minus}\!\left(T_{0}\right)\right)
\end{eqnarray}
and
\begin{eqnarray}
D_{3,1}(1)&=&\kappa\!\left(R_0\right)\left(\kappa\!\left(T,T_{0}\right)+\kappa\!\left(R_{0}\right)\right), \\
D_{3,2}(1)&=&D_{3,1}(1)
\end{eqnarray}
computed by Proposition~\ref{props} and Proposition~\ref{propdk} along with Corollary~\ref{codkb}, respectively, when using the definitions~(\ref{eq:dcs1}) and~(\ref{eq:dcs2}). As we have anticipated, the decoupled term~(\ref{ex:dt3}) only consists of lower-order cumulants available thus far including, for example,  
\begin{equation}
\kappa\!\left(T,T_0\right)=\frac{\dd}{\dd\alpha}\kappa(T)
\end{equation}
obtained with the help of Proposition~\ref{propdab}. 

Upon inserting the necessary lower-order cumulant expressions in the cumulant structure of the second iteration~(\ref{eq:k3i2}), one arrives at an explicit formula of~$\kappa_3(T)$. Converting the formula to that of $S$ via Lemma~\ref{lemmast} while keeping in mind the relation~(\ref{eq:alpha}), we obtain the third cumulant formula of entanglement entropy
\begin{equation}\label{eq:k3}
\kappa_{3}(S)=a_{0}\psi_{2}(mn+1)+a_{1}\psi_{2}(n+1)+a_{2}\psi_1(n)+a_{3}
\end{equation}
with the coefficients listed below. \\

\vspace{-3mm}
\rule{11.1cm}{0.49pt}
\vspace{4pt}
{\vspace{-3pt}\footnotesize\begin{eqnarray*}
a_0&=&1\\
a_1&=&-\frac{m^2+3mn+n^2+1}{(mn+1)(mn+2)} \\
a_2&=&\frac{\left(m^2-1\right)\left(mn-3n^2+1\right)}{n(mn+1)^{2}(mn+2)} \\
a_3&=&-\frac{(m-1)\left(2m^{3}n+3m^{2}n^{2}+2m^{2}+4mn^{3}-15mn^{2}+12mn-2n^{2}-6n+6\right)}{4n^{3}(mn+1)^{2}(mn+2)}
\end{eqnarray*}}
~~~~\!\rule{11.1cm}{0.49pt}
\vspace{9pt}

\noindent It is worth mentioning that for $l=3$ Corollary~\ref{coro2} gives
\begin{equation}
\psi_{2}(mn)-\frac{\kappa\!\left(R_3\right)}{(mn)_{3}}\psi_{2}(n),
\end{equation}
which after applying~(\ref{eq:kR3}) and~(\ref{eq:psisre}) recovers terms involving highest-order polygamma function $\psi_{2}$ in~(\ref{eq:k3}) as expected.

The cumulant formula $\kappa_3(S)$ in~(\ref{eq:k3}) was first derived in~\cite{Wei20} using the existing summation-based methods. Compared to the proof in~\cite{Wei20} of thirty-eight pages, the new methods based on cumulant structures lead to the above proof of less than two pages. Particularly, tasks in~\cite{Wei20} of simplifying up to triple non-trivial summations using about forty tailor-made summation identities while identifying and isolating three anomalies, $\Omega_1$, $\Omega_3$, and $\Omega_{10}$ in Table~\ref{table1}, from the summations turn out to be unnecessary. 

\subsubsection*{Calculation of $\kappa_4$}
Following the implementation of Theorem~\ref{Theorem1} via Algorithm~\ref{alg}, the calculation of fourth cumulant $\kappa_4(T)$ consists of three iterations of computation between decoupled term~(\ref{eq:dT}) and decoupling structure~(\ref{eq:th1}) as
\begin{eqnarray}
\kappa\!\left(T_3,T\right) &=& \delta_2(3)+\frac{\dd}{\dd\alpha}\kappa(T_4), \label{eq:k4i1} \\
\kappa(T_2,T,T) &=& \delta_3(2)+\frac{\dd}{\dd\alpha}\kappa(T_3,T), \label{eq:k4i2} \\
\kappa_4(T) &=& \delta_4(1)+\frac{\dd}{\dd\alpha}\kappa\!\left(T_2,T,T\right). \label{eq:k4i3}
\end{eqnarray}
The first iteration~(\ref{eq:k4i1}) requires the mean 
\begin{equation}\label{eq:mT4}
\kappa(T_4)=\kappa(R_4)\psi_0(m+\alpha)+\Lambda_4
\end{equation}
with
\begin{eqnarray}
\kappa(R_4) &=& m(m+\alpha)(2m+\alpha)\left(7m^2+7m\alpha+\alpha^2+5\right), \label{eq:kR4} \\
\Lambda_4 &=& \frac{m}{60}\big(898m^4+1825m^3\alpha+420m^3+1140m^2\alpha^2+630m^2\alpha \nonumber \\
&& +~\!1310m^2+210m\alpha^3+270m\alpha^2+1595m\alpha+300m+30\alpha^3 \nonumber \\
&& +~\!390\alpha^2+150\alpha+72\big)
\end{eqnarray}
obtained from Lemma~\ref{lemmart}, cf. the structure~(\ref{eq:tl}), where $\kappa\!\left(T_3,T\right)$ becomes available after computing the decoupled term $\delta_2(3)$ by~(\ref{eq:dT}). Taking $\kappa\!\left(T_3,T\right)$ as input, the second iteration~(\ref{eq:k4i2}) produces $\kappa(T_2,T,T)$ by computing $\delta_3(2)$. With the $\kappa(T_2,T,T)$ expression, the last iteration~(\ref{eq:k4i3}) leads to the cumulant structure of $\kappa_4(T)$ by computing the decoupled term $\delta_4(1)$ using~(\ref{eq:dT}) as 
\begin{equation}\label{ex:d41}
\!\!\!\delta_4(1)\!=\!\kappa(R)\left(H_{4,1}(1)+H_{4,2}(1)+H_{4,3}(1)\right)-D_{4,1}(1)-D_{4,2}(1)-D_{4,3}(1) 
\end{equation}
with lower-order cumulants 
\begin{eqnarray}
\!\!\!\!\!\!\!\!\!\!\!H_{4,1}(1)&=&\Big(\kappa\!\left(T,T,T_0\right)-\kappa^{\minus}\!\left(T,T,T_0\right)-2\left(\kappa\!\left(T\right)-\kappa^{\minus}\!\left(T\right)\right) \nonumber \\
&&\times\left(\kappa\!\left(T,T_0\right)-\kappa^{\minus}\!\left(T,T_0\right)\right)-\left(\kappa\!\left(T_0\right)-\kappa^{\minus}\!\left(T_0\right)\right)\left(\kappa_2\!\left(T\right)-\kappa_2^{\minus}\!\left(T\right)\right) \nonumber \\
&&+\left(\kappa\!\left(T_0\right)-\kappa^{\minus}\!\left(T_0\right)\right)\left(\kappa\!\left(T\right)-\kappa^{\minus}\!\left(T\right)\right)^{2}\!\Big)\left(\kappa^{\plus}\!\left(T_0\right)-\kappa\!\left(T_0\right)\right), \label{ex:H431} \\
\!\!\!\!\!\!\!\!\!\!\!H_{4,2}(1)&=&2\big(\kappa\!\left(T,T_0\right)-\kappa^{\minus}\!\left(T,T_0\right)-\left(\kappa\!\left(T_0\right)-\kappa^{\minus}\!\left(T_0\right)\right)\left(\kappa\!\left(T\right)-\kappa^{\minus}\!\left(T\right)\right)\!\big) \nonumber \\
&&\times\big(\kappa^{\plus}\!\left(T,T_0\right)-\kappa\!\left(T,T_0\right)+\left(\kappa^{\plus}\!\left(T_0\right)-\kappa\!\left(T_0\right)\right)\left(\kappa^{\plus}\!\left(T\right)-\kappa\!\left(T\right)\right)\!\big), \label{ex:H432} \\
\!\!\!\!\!\!\!\!\!\!\!H_{4,3}(1)&=&\Big(\kappa^{\plus}\!\left(T,T,T_0\right)-\kappa\!\left(T, T,T_0\right)+2\left(\kappa^{\plus}\!\left(T\right)-\kappa\!\left(T\right)\right) \nonumber \\
&&\times \left(\kappa^{\plus}\!\left(T,T_0\right)-\kappa\!\left(T,T_0\right)\right)+\left(\kappa^{\plus}\!\left(T_0\right)-\kappa\!\left(T_0\right)\right)\left(\kappa_2^{\plus}\!\left(T\right)-\kappa_2\!\left(T\right)\right) \nonumber \\
&&+\left(\kappa^{\plus}\!\left(T_0\right)-\kappa\!\left(T_0\right)\right)\left(\kappa^{\plus}\!\left(T\right)-\kappa\!\left(T\right)\right)^{2}\!\Big)\left(\kappa\!\left(T_0\right)-\kappa^{\minus}\!\left(T_0\right)\right) \label{ex:H433}
\end{eqnarray}
and
\begin{eqnarray}
D_{4,1}(1)&=&2\kappa\!\left(R_0\right)\left(\kappa\!\left(T,T,T_0\right)+\kappa\!\left(T,T_0\right)+\kappa\!\left(R_0\right)\right), \\
D_{4,2}(1)&=&2\left(\kappa\!\left(T,T_0\right)+\kappa\!\left(R_0\right)\right)^2, \\
D_{4,3}(1)&=&D_{4,1}(1)
\end{eqnarray}
computed by Proposition~\ref{props} and Proposition~\ref{propdk} along with Corollary~\ref{codkb}, respectively, when using the definitions~(\ref{eq:dcs1}) and~(\ref{eq:dcs2}). As expected, the decoupled term~(\ref{ex:d41}) involves only lower-order cumulants at hand including, for example,  
\begin{equation}
\kappa\!\left(T,T,T_0\right)=\frac{\dd}{\dd\alpha}\kappa(T,T)
\end{equation}
obtained with the help of Proposition~\ref{propdab}. 

Upon inserting the necessary lower-order cumulant expressions in the cumulant structure of the last iteration~(\ref{eq:k4i3}), one arrives at an explicit formula of~$\kappa_4(T)$. Converting the formula to that of $S$ via Lemma~\ref{lemmast} while keeping in mind the relation~(\ref{eq:alpha}), we obtain the fourth cumulant formula of entanglement entropy
\begin{equation}\label{eq:k4}
\kappa_{4}(S)=b_{0}\psi_{3}(mn+1)+b_{1}\psi_{3}(n+1)+b_{2}\psi_{2}(n)+b_{3}\psi_{1}^{2}(n)+b_{4}\psi_{1}(n)+b_{5}
\end{equation}
with the coefficients listed below. \\

\vspace{-3mm}
\noindent\rule{11.9cm}{0.49pt}
\vspace{2pt}
{\vspace{-6pt}\footnotesize\begin{eqnarray*}
b_0&=&-1\\
b_1&=&\frac{(m+n)\left(m^{2}+5mn+n^{2}+5\right)}{(mn+1)(mn+2)(mn+3)} \nonumber \\
b_2&=&-\frac{\left(m^{2}-1\right)\left(3m^{3}n^{2}-6m^2n^3+9m^2n-12mn^4-6mn^2+6m-20n^3+8n\right)}{n(mn+1)^2(mn+2)^2(mn+3)} \nonumber \\
b_3&=&\frac{6\left(m^2-1\right)\left(n^2-1\right)}{(mn+1)^2(mn+2)(mn+3)} \\
b_4&=&\frac{m^2-1}{n^2(mn+1)^3(mn+2)^2(mn+3)}\left(3m^{4}n^{3}-9m^3n^4+15m^3n^2-6m^2n^4-21m^2n^3\right. \nonumber \\
&&\!\left.~\!+6m^2n^2+24m^2n-36mn^4-18mn^3-4mn^2+18mn+12m-60n^3-12n^2+8n+12\right) \\
b_5&=&-\frac{m-1}{8n^4(mn+1)^3(mn+2)^2(mn+3)}\left(15m^6n^3+20m^{5}n^{4}+27m^5n^3+63m^5n^2+24m^4n^5\right. \nonumber \\
&&~\!-208m^4n^4+257m^4n^3+123m^4n^2+84m^4n+24m^3n^6\!-216m^3n^5+32m^3n^4\!-655m^3n^3 \nonumber \\
&&~\!+765m^3n^2+180m^3n+36m^3-980m^2n^4+200m^2n^3-591m^2n^2+828m^2n+84m^2 \nonumber \\
&&~\!-280mn^4-1220mn^3+576mn^2-180mn+300m-352n^3\!\left.-312n^2+336n-36\right)
\end{eqnarray*}}
\noindent\rule{11.9cm}{0.49pt}
\vspace{-4pt}

\noindent It is worth mentioning that for $l=4$ Corollary~\ref{coro2} gives
\begin{equation}
-\psi_{3}(mn)+\frac{\kappa\!\left(R_4\right)}{(mn)_{4}}\psi_{3}(n),
\end{equation}
which after applying~(\ref{eq:kR4}) and~(\ref{eq:psisre}) indeed recovers terms involving highest-order polygamma function $\psi_{3}$ in~(\ref{eq:k4}).

The cumulant formula $\kappa_4(S)$ in~(\ref{eq:k4}) was first derived in~\cite{HWC21} using the existing summation-based methods. Compared to the twenty-six-page summary of derivation in~\cite{HWC21}, the new methods based on cumulant structures lead to the above derivation of about two pages. In fact, what took some two years for the authors in~\cite{HWC21} to derive the expression~(\ref{eq:k4}) through existing methods now takes a few seconds to generate the same expression on an ordinary laptop when implementing Theorem~\ref{Theorem1} via Algorithm~\ref{alg}. In particular, tasks in~\cite{HWC21} of simplifying up to quadruple summations of non-trivial series using one hundred four tailor-made summation identities are completely avoided. Concurring tasks of identifying and isolating seventeen anomalies, see Table~\ref{table1} for a partial list, from the summations are also not necessary. 

\subsection{Examples of new higher-order cumulants: $\kappa_5$ and $\kappa_6$}\label{sec:3.2}
Using new methods, herein we outline, as examples, the derivation of $\kappa_5$ and $\kappa_6$, formulas of which are unavailable in the literature to our knowledge but can be promptly generated by Algorithm~\ref{alg}.

\subsubsection*{Calculation of $\kappa_5$}
Following the implementation of Theorem~\ref{Theorem1} via Algorithm~\ref{alg}, the calculation of fifth cumulant $\kappa_5(T)$ consists of four iterations of computation between decoupled term~(\ref{eq:dT}) and decoupling structure~(\ref{eq:th1}) as
\begin{eqnarray}
\kappa(T_4,T) &=& \delta_2(4)+\frac{\dd}{\dd\alpha}\kappa(T_{5}), \label{eq:k5i1} \\
\kappa(T_3,T,T) &=& \delta_3(3)+\frac{\dd}{\dd\alpha}\kappa(T_{4},T), \label{eq:k5i2} \\
\kappa(T_2,T,T,T) &=& \delta_4(2)+\frac{\dd}{\dd\alpha}\kappa\!\left(T_3,T,T\right), \label{eq:k5i3} \\  
\kappa_{5}(T) &=& \delta_5(1)+\frac{\dd}{\dd\alpha}\kappa\!\left(T_2,T,T,T\right). \label{eq:k5i4}
\end{eqnarray}
The first iteration~(\ref{eq:k5i1}) requires the mean 
\begin{equation}\label{eq:mT5}
\kappa(T_5)=\kappa(R_5)\psi_0(m+\alpha)+\Lambda_5
\end{equation}
with
\begin{eqnarray}
\kappa(R_5)&=&m(m+\alpha)\big(42m^4+84m^3\alpha+56m^2\alpha^2+70m^2+14m\alpha^3 \nonumber \\
&&+~\!70m\alpha+\alpha^4+15\alpha^2+8\big), \label{eq:kR5} \\
\Lambda_5&=&\frac{m}{30}\big(1417m^5+3621m^4\alpha+630m^4+3230m^3\alpha^2+1260m^3\alpha \nonumber \\
&&+~\!3995m^3+1160m^2\alpha^3+840m^2\alpha^2+6835m^2\alpha+1050m^2 \nonumber \\
&&+~\!135m\alpha^4+210m\alpha^3+3355m\alpha^2+1050m\alpha+1008m \nonumber \\
&&+~\!15\alpha^4+430\alpha^3+225\alpha^2+734\alpha+120\big)
\end{eqnarray}
obtained from Lemma~\ref{lemmart}, cf. the structure~(\ref{eq:tl}), where one arrives at the cumulant structure of $\kappa(T_5)$ by successively computing decoupled terms $\delta_2(4)$, $\delta_3(3)$, $\delta_4(2)$, and $\delta_5(1)$ of the four iterations. We point out that, as discussed in the last paragraph in the proof of Theorem~\ref{Theorem1}, the decoupled term $\delta_l(k)$ when both $l$ and $k$ are greater than or equal to three also requires the decoupling structure~(\ref{eq:th2d}) of joint cumulants up to $\kappa_{l-1}(R_{k-1},T,\dots,T)$, whose decoupled term $\delta_{l}^{(R)}(k)$ is given by~(\ref{eq:dR}). Consequently, among the iterations~(\ref{eq:k5i1})--(\ref{eq:k5i4}), the decoupled term $\delta_3(3)$ in~(\ref{eq:k5i2}) also requires the decoupling structure of $\kappa(R_2,T)$ obtained by using~(\ref{eq:th2d}) as 
\begin{equation}\label{ex:dr22}
\kappa(R_2,T)=\delta_2^{(R)}(2)-\kappa^\prime(T_2)+\frac{\dd}{\dd\alpha}\kappa(R_3).
\end{equation}
In~(\ref{ex:dr22}), $\kappa^\prime(T_2)$ and $\frac{\dd}{\dd\alpha}\kappa(R_3)$ are calculated respectively by Proposition~\ref{propdk} and Proposition~\ref{propdab}, whereas the decoupled term $\delta_2^{(R)}(2)$ is computed by~(\ref{eq:dR}) as
\begin{eqnarray}\label{eq:dtR22}
\!\!\!\!\delta_2^{(R)}(2)&=&\kappa(R)\big(\!\left(\kappa^{\plus}(R)-\kappa(R)\right) \left(\kappa\!\left(T_0\right)-\kappa^{\minus}\!\left(T_0\right)\right)+\left(\kappa\!\left(R_0\right)-\kappa^{\minus}\!\left(R_0\right)\right) \nonumber \\
&&\times\left(\kappa^{\plus}(T)-\kappa(T)\right)+\left(\kappa^{\plus}\!\left(R_0\right)-\kappa\!\left(R_0\right)\right)\left(\kappa(T)-\kappa^{\minus}(T)\right) \nonumber \\
&&+\left(\kappa(R)-\kappa^{\minus}(R)\right)\left(\kappa^{\plus}\!\left(T_0\right)-\kappa\!\left(T_0\right)\right)\!\big)-2\kappa\!\left(R_0\right)\kappa(R).
\end{eqnarray}

Upon inserting the needed lower-order cumulant expressions in the last iteration~(\ref{eq:k5i4}), one arrives at an explicit formula of~$\kappa_5(T)$. Converting the formula to that of $S$ via Lemma~\ref{lemmast} while keeping in mind the relation~(\ref{eq:alpha}), we obtain the fifth cumulant formula of entanglement entropy 
\begin{eqnarray}\label{eq:k5}
\kappa_5(S) &=& c_0\psi_4(mn+1)+c_1\psi_4(n+1)+c_2\psi_3(n)+c_3\psi_1(n)\psi_2(n) \nonumber \\
&& +~\!c_4\psi_2(n)+c_5\psi_1^2(n)+c_6\psi_1(n)+c_7
\end{eqnarray}
with the coefficients listed below. \\

\vspace{-3mm}
\noindent\rule{11.9cm}{0.49pt}
{\vspace{-4pt}\footnotesize\begin{eqnarray*}
\!\!\!c_0&=&1\\
\!\!\!c_1&=&-\frac{m^4+10 m^3 n+20 m^2 n^2+15 m^2+10 m n^3+40 m n+n^4+15 n^2+8}{(mn+1)(mn+2)(mn+3)(mn+4)}\\
\!\!\!c_2&=&\frac{2\left(m^2-1\right)}{n(mn+1)^2(mn+2)^2(mn+3)^2(mn+4)}\left(3m^5n^3+18m^4n^2-30m^3n^5+23m^3n^3\right.\nonumber\\
&&+~\!33m^3n-15m^2n^6-155m^2n^4+113m^2n^2+18m^2-65mn^5-240mn^3+153mn\nonumber\\
&&\!\left.-~\!65n^4-115n^2+48\right)\\
\!\!\!c_3&=&-\frac{60\left(m^2-1\right) (m+n)\left(n^2-1\right) }{(mn+1)^2(mn+2)(mn+3)(mn+4)}\\
\!\!\!c_4&=&-\frac{5\left(m^2-1\right)}{n^2(mn+1)^3(mn+2)^3(mn+3)^2(mn+4)}\left(3m^7n^5-9m^6n^6+30m^6n^4-3m^5n^7\right.\nonumber\\
&&-~\!6m^5n^6-57m^5n^5+6m^5n^4+117m^5n^3-6m^4n^7-60m^4n^6-42m^4n^5-99m^4n^4\nonumber\\
&&+~\!48m^4n^3+222m^4n^2-60m^3n^7-48m^3n^6-273m^3n^5-90m^3n^4+27m^3n^3\nonumber\\
&&+~\!138m^3n^2+204m^3n-360m^2n^6-138m^2n^5-420m^2n^4-30m^2n^3+198m^2n^2\nonumber\\
&&+~\!168m^2n+72m^2-700mn^5-168mn^4-116mn^3+96mn^2+120mn+72m-448n^4\nonumber\\
&&\!\left.-~\!72n^3+136n^2+72n\right)\\
\!\!\!c_5&=&-\frac{30\left(m^2-1\right)}{n(mn+1)^3(mn+2)^2(mn+3)^2(mn+4)}
\left(2m^4n^3+2m^3n^6-6m^3n^4+12m^3n^2\right.\nonumber\\
&&\!\left.+~\!3m^2n^5-27m^2n^3+22m^2n-19mn^4-25mn^2+12m-32n^3+8n\right)\\
\!\!\!c_6&=&\frac{5\left(m^2-1\right)}{n^3(mn+1)^4(mn+2)^3(mn+3)^2(mn+4)}\left(3m^8n^6-9m^7n^7+39m^7n^5-12m^6n^7\right.\nonumber\\
&&-~\!81m^6n^6+24m^6n^5+201m^6n^4-75m^5n^7-132m^5n^6-252m^5n^5+216m^5n^4\nonumber\\
&&+~\!525m^5n^3-54m^4n^7-552m^4n^6-534m^4n^5-225m^4n^4+744m^4n^3+732m^4n^2\nonumber\\
&&-~\!180m^3n^7-408m^3n^6-1467m^3n^5-948m^3n^4+387m^3n^3+1224m^3n^2+516m^3n\nonumber\\
&&-~\!1080m^2n^6-1110m^2n^5-1670m^2n^4-594m^2n^3+972m^2n^2+960m^2n+144m^2\nonumber\\
&&-~\!2100mn^5-1284mn^4-688mn^3+180mn^2+684mn+288m-1344n^4-528n^3\nonumber\\
&&\!\left.-~\!8n^2+240n+144\right)\\
\!\!\!c_7&=&-\frac{m-1}{8n^5(mn+1)^4(mn+2)^3(mn+3)^2(mn+4)}\left(84m^{10}n^6+105m^9n^7+312m^9n^6\right.\nonumber\\
&&+~\!900m^9n^5+120m^8n^8-1395m^8n^7+2832m^8n^6+3480m^8n^5+3900m^8n^4+120m^7n^9\nonumber\\
&&-~\!3360m^7n^8+1215m^7n^7-11688m^7n^6+22185m^7n^5+15720m^7n^4+8748m^7n^3\nonumber\\
&&+~\!96m^6n^{10}-1560m^6n^9-960m^6n^8-38505m^6n^7+11972m^6n^6-35595m^6n^5\nonumber\\
&&+~\!81930m^6n^4+36744m^6n^3+10704m^6n^2+240m^5n^9-18840m^5n^8-18240m^5n^7\nonumber\\
&&-~\!173680m^5n^6+71035m^5n^5-42870m^5n^4+164184m^5n^3+46752m^5n^2+6768m^5n\nonumber\\
&&-~\!4560m^4n^8-93420m^4n^7-81280m^4n^6-394985m^4n^5+216890m^4n^4+624m^4n^3\nonumber\\
&&+~\!183312m^4n^2+30624m^4n+1728m^4-31728m^3n^7-238280m^3n^6-153320m^3n^5\nonumber\\
&&-~\!480010m^3n^4+351564m^3n^3+46512m^3n^2+106944m^3n+8064m^3-84784m^2n^6\nonumber\\
&&-~\!319780m^2n^5-120640m^2n^4-302040m^2n^3+300192m^2n^2+36384m^2n+25344m^2\nonumber\\
&&-~\!104768mn^5-203240mn^4-18480mn^3-87120mn^2+122064mn+8064m-49888n^4\nonumber\\
&&-~\!41280n^3+\!\left.11520n^2-8640n+16704\right)
\end{eqnarray*}}
\noindent\rule{11.9cm}{0.49pt}
\vspace{0pt}

\noindent It is worth mentioning that for $l=5$ Corollary~\ref{coro2} gives
\begin{equation}
\psi_{4}(mn)-\frac{\kappa\!\left(R_5\right)}{(mn)_{5}}\psi_{4}(n),
\end{equation}
which after applying~(\ref{eq:kR5}) and~(\ref{eq:psisre}) indeed recovers terms involving highest-order polygamma function $\psi_{4}$ in~(\ref{eq:k5}).

\subsubsection*{Calculation of $\kappa_6$}
Following the implementation of Theorem~\ref{Theorem1} via Algorithm~\ref{alg}, the calculation of sixth cumulant $\kappa_6(T)$ consists of five iterations of computation between decoupled term~(\ref{eq:dT}) and decoupling structure~(\ref{eq:th1}) as
\begin{eqnarray}
\kappa\!\left(T_5,T\right) &=& \delta_2(5)+\frac{\dd}{\dd\alpha}\kappa\!\left(T_6\right), \label{eq:k6i1} \\
\kappa\!\left(T_4,T,T\right) &=& \delta_3(4)+\frac{\dd}{\dd\alpha}\kappa\!\left(T_5,T\right), \label{eq:k6i2} \\
\kappa\!\left(T_3,T,T,T\right) &=& \delta_4(3)+\frac{\dd}{\dd\alpha}\kappa\!\left(T_4,T,T\right), \label{eq:k6i3} \\
\kappa\!\left(T_2,T,T,T,T\right) &=& \delta_5(2)+\frac{\dd}{\dd\alpha}\kappa\!\left(T_3,T,T,T\right), \label{eq:k6i4}  \\
\kappa_6(T) &=& \delta_6(1)+\frac{\dd}{\dd\alpha}\kappa\!\left(T_2,T,T,T,T\right). \label{eq:k6i5}
\end{eqnarray}
The first iteration~(\ref{eq:k6i1}) requires the mean 
\begin{equation}\label{eq:mT6}
\kappa(T_6)=\kappa(R_6)\psi_0(m+\alpha)+\Lambda_6
\end{equation}
with
\begin{eqnarray}
\kappa(R_6)&=&m(m+\alpha)(2m+\alpha)\big(66m^4+132m^3\alpha+84m^2\alpha^2+210m^2 \nonumber \\
&&+~\!18m\alpha^3+210m\alpha+\alpha^4+35\alpha^2+84\big), \label{eq:kR6} \\
\Lambda_6&=&\frac{m}{210}\big(32254m^6+99029m^5\alpha+13860m^5+114135m^4\alpha^2 \nonumber \\
&&+~\!34650m^4\alpha+155890m^4+60550m^3\alpha^3+31500m^3\alpha^2 \nonumber \\
&&+~\!343315m^3\alpha+44100m^3+14350m^2\alpha^4+12600m^2\alpha^3 \nonumber \\
&&+~\!252875m^2\alpha^2+66150m^2\alpha+103096m^2+1155m\alpha^5 \nonumber \\
&&+~\!2100m\alpha^4+70175m\alpha^3+29400m\alpha^2+130116m\alpha \nonumber \\
&&+~\!17640m+105\alpha^5+5600\alpha^4+3675\alpha^3+35140\alpha^2 \nonumber \\
&&+~\!8820\alpha+3600\big)
\end{eqnarray}
obtained from Lemma~\ref{lemmart}, cf. the structure~(\ref{eq:tl}), where one arrives at the cumulant structure of $\kappa(T_6)$ by successively computing decoupled terms $\delta_2(5)$, $\delta_3(4)$, $\delta_4(3)$, $\delta_5(2)$, and $\delta_6(1)$ of the iterations from~(\ref{eq:k6i1}) to~(\ref{eq:k6i5}). There, the decoupled terms $\delta_3(4)$ in~(\ref{eq:k6i2}) and $\delta_4(3)$ in~(\ref{eq:k6i3}) require respectively the decoupling structures of
\begin{equation}\label{eq:dr23}
\kappa(R_3,T)=\delta_2^{(R)}(3)-\kappa^\prime(T_3)+\frac{\dd}{\dd\alpha}\kappa(R_{4})
\end{equation}
and 
\begin{eqnarray}\label{eq:dr32}
\kappa(R_2,T,T)=\delta_3^{(R)}(2)-\kappa^\prime(T_2,T)+\frac{\dd}{\dd\alpha}\kappa(R_{3},T).
\end{eqnarray}
The corresponding decoupled terms $\delta_2^{(R)}(3)$ and $\delta_3^{(R)}(2)$ are computed by using~(\ref{eq:dR}) as
\begin{eqnarray}\label{eq:dr23cs}
\!\!\delta_2^{(R)}(3)&=&\kappa(R)\big(\!\left(\kappa^\plus\!\left(R_2\right)-\kappa\!\left(R_2\right)\right)\left(\kappa\!\left(T_0\right)-\kappa^\minus\!\left(T_0\right)\right)+\left(\kappa(R_2)-\kappa^\minus(R_2)\right)\nonumber\\
\!\!&&\times\left(\kappa^\plus(T_0)-\kappa(T_0)\right)+\left(\kappa^\plus(R)-\kappa(R)\right)\left(\kappa(T)-\kappa^\minus(T)\right)\nonumber\\
\!\!&&+\left(\kappa\!\left(R\right)-\kappa^\minus\!\left(R\right)\right)\left(\kappa^\plus\!\left(T\right)-\kappa\!\left(T\right)\right)+\left(\kappa^\plus\!\left(R_0\right)-\kappa\!\left(R_0\right)\right)\nonumber\\
\!\!&&\times\left(\kappa\!\left(T_2\right)-\kappa^\minus\!\left(T_2\right)\right)+\left(\kappa\!\left(R_0\right)-\kappa^\minus\!\left(R_0\right)\right) \left(\kappa^\plus\!\left(T_2\right)-\kappa\!\left(T_2\right)\right)\nonumber\\
\!\!&&-~\!2\kappa(R)-2\kappa(T)\big)-4 \kappa\!\left(R_0\right)\kappa\!\left(R_2\right),
\end{eqnarray}
\begin{eqnarray}\label{eq:dr32cs}
\delta_3^{(R)}(2)&=&\kappa(R)\Big(\!\left(\kappa^\plus(R)-\kappa (R)\right) \left(\kappa\!\left(T,T_0\right)-\kappa^\minus\!\left(T,T_0\right)-\left(\kappa (T)-\kappa^\minus(T)\right) \right.\nonumber\\
&&\times\left.\!\left(\kappa\!\left(T_0\right)-\kappa^\minus\!\left(T_0\right)\right)\right)+\left(\kappa_2(T)-\kappa_2^\minus(T)-\left(\kappa(T)-\kappa^\minus(T)\right)^2\right)\nonumber\\
&&\times\left(\kappa^\plus\!\left(R_0\right)-\kappa\!\left(R_0\right)\right)+\left(\kappa^\plus\!\left(T_0\right)-\kappa\!\left(T_0\right)\right)\left(\kappa(R)-\kappa^\minus(T,R)\right.\nonumber\\
&&\!\left.-\left(\kappa(R)-\kappa^\minus(R)\right)\left(\kappa(T)-\kappa^\minus(T)\right)+\kappa(T)\right)+\left(\kappa\!\left(T_0\right)-\kappa^\minus\!\left(T_0\right)\right)\nonumber\\
&&\times\left(\kappa^\plus(T,R)-\kappa (R)-\kappa (T)+\left(\kappa^\plus(R)-\kappa(R)\right)\left(\kappa^\plus(T)-\kappa(T)\right)\right)\nonumber\\
&&+\left(\kappa\!\left(R_0\right)-\kappa^\minus\!\left(R_0\right)\right)\left(\kappa_2^\plus(T)-\kappa_2(T)+\left(\kappa^\plus(T)-\kappa (T)\right)^2\right)\nonumber\\
&&+\left(\kappa^\plus\!\left(T,T_0\right)-\kappa\!\left(T,T_0\right)+\left(\kappa^\plus(T_0)-\kappa(T_0)\right)\left(\kappa^\plus\!\left(T\right)-\kappa\!\left(T\right)\right)\right)\nonumber\\
&&\times\left(\kappa(R)-\kappa^\minus(R)\right)-2\left(\kappa\!\left(T,T_0\right)+\kappa\!\left(R_0\right)\right)\!\Big)\nonumber\\
&&-~\!2\kappa\!\left(R_0\right)(3\kappa(R)+2\kappa(T)),
\end{eqnarray}
which only involve cumulants of lower order, relative to $\kappa(R_3,T)$ and $\kappa(R_2,T,T)$, respectively, available thus far including, for example,  
\begin{equation}
\kappa^{\substack{\normalfont\plus\vspace{-0.24em} \\ \normalfont\minus}}\!\left(T,R\right) =
\frac{\dd}{\dd\beta}\kappa^{\substack{\normalfont\plus\vspace{-0.24em} \\ \normalfont\minus}}(T)
\end{equation}
obtained with the help of~(\ref{eq:matdb}). 

Upon inserting the necessary lower-order cumulant expressions in the cumulant structure of the last iteration~(\ref{eq:k6i5}), one arrives at an explicit formula of~$\kappa_6(T)$. Converting the formula to that of $S$ via Lemma~\ref{lemmast} while keeping in mind the relation~(\ref{eq:alpha}), we obtain the sixth cumulant formula of entanglement entropy
\begin{eqnarray}\label{eq:k6}
\kappa_6(S)&=&d_{0}\psi_5(mn+1)+d_{1}\psi_5(n+1)+d_{2}\psi_4(n)+d_{3}\psi_1(n)\psi_3(n) \nonumber \\
&&+~\!d_{4}\psi_3(n)+d_{5}\psi_{2}^{2}(n)+d_{6}\psi_1(n)\psi_2(n)+d_{7}\psi_2(n) \nonumber \\
&&+~\!d_{8}\psi_{1}^{3}(n)+d_{9}\psi_{1}^{2}(n)+d_{10}\psi_1(n)+d_{11} 
\end{eqnarray}
with the coefficients listed below. \\

\vspace{-3mm}
\noindent\rule{11.9cm}{0.49pt}
{\vspace{-4pt}\footnotesize
\begin{eqnarray*}
d_0&=&-1\\
d_1&=&\frac{1}{(mn+1)(mn+2)(mn+3)(mn+4)(mn+5)}\Big(m^5+15m^4n+50m^3n^2+35m^3\nonumber\\
&&+~\!50m^2n^3+175m^2n+15mn^4+175mn^2+84m+n^5+35n^3+84n\Big)\nonumber\\
d_2&=&-\frac{2\left(m^2-1\right)}{n(mn+1)^2(mn+2)^2(mn+3)^2(mn+4)^2(mn+5)}\Big(5m^7n^4+15m^6n^5+50m^6n^3\nonumber\\
&&-~\!75m^5n^6+235m^5n^4+175m^5n^2-125m^4n^7-685m^4n^5+1315m^4n^3+250m^4n\nonumber\\
&&-~\!30m^3n^8-1250m^3n^6-2095m^3n^4+3275m^3n^2+120m^3-240m^2n^7-4537m^2n^5\nonumber\\
&&-~\!2135m^2n^3+3572m^2n-600mn^6-7150mn^4+670mn^2+1320m-462n^5\nonumber\\
&&-~\!4170n^3+1752n\Big)\nonumber\\
d_3&=&\frac{60\left(m^2-1\right)\left(n^2-1\right)\left(3m^2+8mn+3n^2+8\right)}{(mn+1)^2(mn+2)(mn+3)(mn+4)(mn+5)}\nonumber\\
d_4&=&\frac{m^2-1}{n^2(mn+1)^3(mn+2)^3(mn+3)^3(mn+4)^2(mn+5)}\Big(45m^{10}n^7-90m^9n^8+765m^9n^6\nonumber\\
&&-~\!270m^8n^9-90m^8n^8-795m^8n^7+90m^8n^6+5445m^8n^5+105m^7n^{10}-240m^7n^9\nonumber\\
&&-~\!4745m^7n^8-1110m^7n^7+15m^7n^6+1350m^7n^5+21015m^7n^4-90m^6n^{10}-825m^6n^9\nonumber\\
&&-~\!3750m^6n^8-33155m^6n^7-4350m^6n^6+24135m^6n^5+8190m^6n^4+47430m^6n^3\nonumber\\
&&-~\!1350m^5n^{10}-1350m^5n^9-21495m^5n^8-24090m^5n^7-117155m^5n^6-210m^5n^5\nonumber\\
&&+~\!116439m^5n^4+25650m^5n^3+62460m^5n^2-16650m^4n^9-8190m^4n^8-130725m^4n^7\nonumber\\
&&-~\!82050m^4n^6-215735m^4n^5+46680m^4n^4+254724m^4n^3+43560m^4n^2+44280m^4n\nonumber\\
&&-~\!79950m^3n^8-25650m^3n^7-374070m^3n^6-158910m^3n^5-176636m^3n^4\nonumber\\
&&+~\!146760m^3n^3+289284m^3n^2+37800m^3n+12960m^3-186870m^2n^7-43560m^2n^6\nonumber\\
&&-~\!561690m^2n^5-173400m^2n^4+544m^2n^3+204000m^2n^2+162144m^2n+12960m^2\nonumber\\
&&-~\!212820mn^6-37800mn^5-427500mn^4-97560mn^3+75264mn^2+135360mn\nonumber\\
&&+~\!34560m-94920n^5-12960n^4-129720n^3-21600n^2+21024n+34560\Big)\nonumber\\
d_5&=&\frac{30\left(m^2-1\right)\left(n^2-1\right)\left(6m^3n+13m^2n^2+14m^2+6mn^3+33mn+14n^2+4\right)}{(mn+1)^2(mn+2)^2(mn+3)(mn+4)(mn+5)}\nonumber\\
d_6&=&-\frac{60\left(m^2-1\right)}{n(mn+1)^3(mn+2)^2(mn+3)^2(mn+4)^2(mn+5)}\Big(3m^6n^6-15m^6n^4+65m^5n^5\nonumber\\
&&-~\!21m^5n^7-152m^5n^3-12m^4n^8-105m^4n^6+382m^4n^4-541m^4n^2-42m^3n^7\nonumber\\
&&+~\!115m^3n^5+775m^3n^3-788m^3n+228m^2n^6+1037m^2n^4+271m^2n^2-384m^2\nonumber\\
&&+~\!1110mn^5+754mn^3-376mn+1140n^4-660n^2+96\Big)\nonumber\\
d_7&=&-\frac{15\left(m^2-1\right)}{n^3(mn+1)^4(mn+2)^4(mn+3)^3(mn+4)^2(mn+5)}\Big(7m^{12}n^9-24m^{11}n^{10}\nonumber\\
&&+~\!154m^{11}n^8+2m^{10}n^{11}-30m^{10}n^{10}-403m^{10}n^9+54m^{10}n^8+1474m^{10}n^7-10m^9n^{11}\nonumber\\
&&-~\!214m^9n^{10}-566m^9n^9-2734m^9n^8+984m^9n^7+8044m^9n^6-110m^8n^{11}-394m^8n^{10}\nonumber\\
&&-~\!3592m^8n^9-4418m^8n^8-9070m^8n^7+7644m^8n^6+27535m^8n^5-164m^7n^{11}\nonumber\\
&&-~\!2384m^7n^{10}-4784m^7n^9-23464m^7n^8-18140m^7n^7-11392m^7n^6+33024m^7n^5\nonumber\\
&&+~\!61186m^7n^4-840m^6n^{11}-2672m^6n^{10}-20480m^6n^9-28268m^6n^8-79498m^6n^7\nonumber\\
&&-~\!39938m^6n^6+19481m^6n^5+86646m^6n^4+88056m^6n^3-11760m^5n^{10}-18224m^5n^9\nonumber\\
&&-~\!91868m^5n^8-93362m^5n^7-146242m^5n^6-35630m^5n^5+98750m^5n^4+141096m^5n^3\nonumber\\
&&+~\!78936m^5n^2-67200m^4n^9-67400m^4n^8-233810m^4n^7-177386m^4n^6-126600m^4n^5\nonumber\\
&&+~\!31162m^4n^4+165816m^4n^3+138936m^4n^2+39888m^4n-201040m^3n^8-145820m^3n^7\nonumber\\
&&-~\!335860m^3n^6-181156m^3n^5+1920m^3n^4+107904m^3n^3+141192m^3n^2+75456m^3n\nonumber\\
&&+~\!8640m^3-332920m^2n^7-184232m^2n^6-242720m^2n^5-68992m^2n^4+89448m^2n^3\nonumber\\
&&+~\!98472m^2n^2+58608m^2n+17280m^2-290240mn^6-125520mn^5-49504mn^4\nonumber\\
&&+~\!25872mn^3+56400mn^2+31680mn+8640m-104384n^5-35424n^4+19808n^3\nonumber\\
&&+~\!21600n^2+8640n\Big)\nonumber\\
d_8&=&\frac{120\left(m^2-1\right)\left(n^2-1\right)\left(m^2n+mn^2-5m-5n\right)}{(mn+1)^3(mn+2)(mn+3)(mn+4)(mn+5)}\nonumber\\
d_9&=&-\frac{30\left(m^2-1\right)}{n^2(mn+1)^4(mn+2)^3(mn+3)^3(mn+4)^2(mn+5)}\Big(3m^9n^9+21m^9n^7+6m^8n^{10}\nonumber\\
&&+~\!6m^8n^9-54m^8n^8-6m^8n^7+356m^8n^6-6m^7n^{11}+6m^7n^{10}+153m^7n^9+54m^7n^8\nonumber\\
&&-~\!1143m^7n^7-60m^7n^6+2524m^7n^5+60m^6n^{10}+60m^6n^9+1116m^6n^8+36m^6n^7\nonumber\\
&&-~\!7226m^6n^6-96m^6n^5+9690m^6n^4+1080m^5n^9+96m^5n^8+2640m^5n^7-1116m^5n^6\nonumber\\
&&-~\!21775m^5n^5+1020m^5n^4+21719m^5n^3+4056m^4n^8-1020m^4n^7-3048m^4n^6\nonumber\\
&&-~\!4626m^4n^5-33078m^4n^4+5646m^4n^3+28354m^4n^2+930m^3n^7-5646m^3n^6\nonumber\\
&&-~\!24333m^3n^5-6354m^3n^4-21869m^3n^3+12000m^3n^2+19896m^3n-23724m^2n^6\nonumber\\
&&-~\!12000m^2n^5-37218m^2n^4+264m^2n^3-1786m^2n^2+11736m^2n+5760m^2\nonumber\\
&&-~\!48300mn^5-11736mn^4-13956mn^3+7416mn^2+1200mn+4320m-29280n^4\nonumber\\
&&-~\!4320n^3+6240n^2+4320n-1440\Big)\nonumber\\
d_{10}&=&\frac{15\left(m^2-1\right)}{n^4(mn+1)^5(mn+2)^4(mn+3)^3(mn+4)^2(mn+5)}\Big(7m^{13}n^{10}-21m^{12}n^{11}\nonumber\\
&&+~\!182m^{12}n^9-42m^{11}n^{11}-441m^{11}n^{10}+138m^{11}n^9+2036m^{11}n^8-273m^{10}n^{11}\nonumber\\
&&-~\!1046m^{10}n^{10}-3750m^{10}n^9+2670m^{10}n^8+12956m^{10}n^7-366m^9n^{11}-5116m^9n^{10}\nonumber\\
&&-~\!10534m^9n^9-15816m^9n^8+22380m^9n^7+52067m^9n^6-1416m^8n^{11}-7018m^8n^{10}\nonumber\\
&&-~\!41204m^8n^9-56386m^8n^8-27285m^8n^7+106620m^8n^6+138302m^8n^5-1032m^7n^{11}\nonumber\\
&&-~\!22380m^7n^{10}-57452m^7n^9-185672m^7n^8-172822m^7n^7+40683m^7n^6\nonumber\\
&&+~\!317994m^7n^5+246154m^7n^4-2520m^6n^{11}-16296m^6n^{10}-148848m^6n^9\nonumber\\
&&-~\!261644m^6n^8-508349m^6n^7-289042m^6n^6+321432m^6n^5+615390m^6n^4\nonumber\\
&&+~\!290064m^6n^3-35280m^5n^{10}-107712m^5n^9-541680m^5n^8-722126m^5n^7\nonumber\\
&&-~\!850724m^5n^6-162490m^5n^5+767406m^5n^4+772320m^5n^3+216696m^5n^2\nonumber\\
&&-~\!201600m^4n^9-386208m^4n^8-1171112m^4n^7-1227194m^4n^6-802694m^4n^5\nonumber\\
&&+~\!289858m^4n^4+1001016m^4n^3+605880m^4n^2+92736m^4n-603120m^3n^8\nonumber\\
&&-~\!810792m^3n^7-1521316m^3n^6-1230632m^3n^5-278824m^3n^4+644256m^3n^3\nonumber\\
&&+~\!754728m^3n^2+269568m^3n+17280m^3-998760m^2n^7-995544m^2n^6-1133232m^2n^5\nonumber\\
&&-~\!624304m^2n^4+173880m^2n^3+513960m^2n^2+307008m^2n+51840m^2-870720mn^6\nonumber\\
&&-~\!660720mn^5-421424mn^4-70224mn^3+187776mn^2+176256mn+51840m\nonumber\\
&&-~\!313152n^5-182208n^4-52192n^3+39360n^2+46080n+17280\Big)\nonumber\\
d_{11}&=&-\frac{3(m-1)}{4n^6(mn+1)^5(mn+2)^4(mn+3)^3(mn+4)^2(mn+5)}\Big(105m^{15}n^{10}+126m^{14}n^{11}\nonumber\\
&&+~\!575m^{14}n^{10}+2270m^{14}n^9+140m^{13}n^{12}-1644m^{13}n^{11}+5930m^{13}n^{10}+12690m^{13}n^9\nonumber\\
&&+~\!21640m^{13}n^8+140m^{12}n^{13}-7300m^{12}n^{12}+3886m^{12}n^{11}-26670m^{12}n^{10}\nonumber\\
&&+~\!92820m^{12}n^9+123480m^{12}n^8+119740m^{12}n^7+120m^{11}n^{14}-7440m^{11}n^{13}\nonumber\\
&&-~\!2120m^{11}n^{12}-165774m^{11}n^{11}+70105m^{11}n^{10}-163380m^{11}n^9+758472m^{11}n^8\nonumber\\
&&+~\!697060m^{11}n^7+425605m^{11}n^6+80m^{10}n^{15}-2040m^{10}n^{14}-3240m^{10}n^{13}\nonumber\\
&&-~\!176540m^{10}n^{12}-79884m^{10}n^{11}-1635305m^{10}n^{10}+769750m^{10}n^9-353568m^{10}n^8\nonumber\\
&&+~\!3793246m^{10}n^7+2525635m^{10}n^6+1014550m^{10}n^5+640m^9n^{14}-49140m^9n^{13}\nonumber\\
&&-~\!110320m^9n^{12}-1857198m^9n^{11}-795620m^9n^{10}-9227510m^9n^9+5167552m^9n^8\nonumber\\
&&+~\!945276m^9n^7+12480082m^9n^6+6130330m^9n^5+1641010m^9n^4-8680m^8n^{13}\nonumber\\
&&-~\!531860m^8n^{12}-1237768m^8n^{11}-11358200m^8n^{10}-3932060m^8n^9-32956288m^8n^8\nonumber\\
&&+~\!22272946m^8n^7+8395242m^8n^6+27895072m^8n^5+10082750m^8n^4+1776560m^8n^3\nonumber\\
&&-~\!168840m^7n^{12}-3402448m^7n^{11}-7554960m^7n^{10}-44561560m^7n^9-10560268m^7n^8\nonumber\\
&&-~\!77694994m^7n^7+64008017m^7n^6+25850532m^7n^5+42638060m^7n^4+11082640m^7n^3\nonumber\\
&&+~\!1230840m^7n^2-1256304m^6n^{11}-14187880m^6n^{10}-28548440m^6n^9-116889096m^6n^8\nonumber\\
&&-~\!13032404m^6n^7-121969113m^6n^6+124939362m^6n^5+45708660m^6n^4+43840240m^6n^3\nonumber\\
&&+~\!7783240m^6n^2+492480m^6n-5449680m^5n^{10}-40013060m^5n^9-69918016m^5n^8\nonumber\\
&&-~\!207369398m^5n^7+5930952m^5n^6-124913458m^5n^5+165517530m^5n^4+50063040m^5n^3\nonumber\\
&&+~\!28941520m^5n^2+3151680m^5n+86400m^5-15166600m^4n^9-76563156m^4n^8\nonumber\\
&&-~\!111628488m^4n^7-246008276m^4n^6+46478552m^4n^5-78485210m^4n^4\nonumber\\
&&+~\!145533360m^4n^3+33459120m^4n^2+11057280m^4n+558720m^4-27565928m^3n^8\nonumber\\
&&-~\!96949288m^3n^7-113053336m^3n^6-188138076m^3n^5+72464840m^3n^4-25967120m^3n^3\nonumber\\
&&+~\!80520120m^3n^2+12458880m^3n+1854720m^3-31740464m^2n^7-76411416m^2n^6\nonumber\\
&&-~\!67560816m^2n^5-86031480m^2n^4+56248560m^2n^3-2561400m^2n^2+25030080m^2n\nonumber\\
&&+~\!1969920m^2-21051968mn^6-32850416mn^5-20208480mn^4-20171520mn^3\nonumber\\
&&+~\!21997440mn^2+267840mn+3265920m-6132928n^5-5464320n^4-1900800n^3\nonumber\\
&&-~\!1641600n^2+3352320n-86400\Big)
\end{eqnarray*}}
\noindent\rule{11.9cm}{0.49pt}
\vspace{0pt}

\noindent It is worth mentioning that for $l=6$ Corollary~\ref{coro2} gives
\begin{equation}
-\psi_{5}(mn)+\frac{\kappa\!\left(R_6\right)}{(mn)_{6}}\psi_{5}(n),
\end{equation}
which after applying~(\ref{eq:kR6}) and~(\ref{eq:psisre}) indeed recovers terms involving highest-order polygamma function $\psi_{5}$ in~(\ref{eq:k6}).

\section{Conclusion}\label{sec4}
In this work, we propose a new framework to derive exact yet explicit cumulant formulas of any order of entanglement entropy over Hilbert-Schmidt ensemble. The framework consists of a set of cumulant structural results that enable decoupling of cumulants into lower-order joint cumulants involving ancillary linear statistics. The new decoupling procedure circumvents the need of existing methods to simplify nested summations that becomes prohibitively tedious as the order of cumulant increases. Future works include finding cumulant structures of entropy over other major generic state models such as Bures-Hall ensemble and fermionic Gaussian ensemble.
 
\backmatter
\bmhead{Acknowledgment} The work of Lu Wei was supported by the U.S. National Science Foundation (2306968) and the U.S. Department of Energy (DE-SC0024631).

\appendix
\section*{Appendices}\label{sec:app}
\addcontentsline{toc}{section}{\nameref{sec:app}}

\section{Proof of Lemma~\ref{lemmast}}\label{appst}
\begin{proof}
The change of variables
\begin{equation}\label{eq:cvst}
\lambda_i=\frac{x_i}{r},
\end{equation}
for $i=1,\dots,m$, leads to factorization of the Wishart-Laguerre ensemble~(\ref{eq:we}) into product of the density of trace 
\begin{equation}
r=\sum_{i=1}^{m}x_i
\end{equation}
and the Hilbert-Schmidt ensemble~(\ref{eq:fte}) as~\cite{Page93,Wei17}
\begin{equation}\label{eq:factorize}
g(\bm{x})\prod_{i=1}^{m}\dd x_i=h_{mn}(r)f(\bm{\lambda})\dd r\prod_{i=1}^m\dd\lambda_i,
\end{equation}
where
\begin{equation}\label{eq:fc}
h_{m n}(r)=\frac{1}{\Gamma(mn)}\ee^{-r}r^{mn-1},
\end{equation}
$r\in[0,\infty)$, is the density of $r$. The change of variables~(\ref{eq:cvst}) also leads to the relation between the two linear statistics~(\ref{eq:vN}) and~(\ref{eq:von}),
\begin{eqnarray}
S&=&r^{-1}\left(r\ln r-T\right), \label{eq:TtoS} \\
T&=&r\left(\ln r-S\right). \label{eq:StoT}
\end{eqnarray}

The $l$-th moment of $S$ can now be converted into the $l$-th moment of $T$ as
\begin{eqnarray}
\mathbb{E}\!\left[S^l\right]&=&\int_{\bm{\lambda}}S^lf(\bm{\lambda})\prod_{i=1}^{m}\dd\lambda_i\int_{r}h_{mn+l}(r)\dd r \\
&=&\sum_{i=0}^{l-1}\sum_{j=0}^{i}(-1)^{i+j}\binom{l}{i}\binom{i}{j}\int_{\bm{\lambda}}S^{j}f(\bm{\lambda})\prod_{i=1}^{m}\dd\lambda_i\int_{r}\ln^{l-j}r~\!h_{mn+l}(r)\dd r \nonumber \\
&&+(-1)^l\frac{\Gamma(mn)}{\Gamma(mn+l)}\mathbb{E}\!\left[T^l\right] \label{eq:sts3} \\
&=&\sum_{j=0}^{l-1}(-1)^{j+l+1}\binom{l}{j}\mathbb{E}\!\left[S^j\right]\!\left.\frac{\dd^{l-j}}{\dd a^{l-j}}\frac{\Gamma(a)}{\Gamma(mn+l)}\right\rvert_{a=mn+l} \nonumber \\
&&+(-1)^l\frac{\Gamma(mn)}{\Gamma(mn+l)}\mathbb{E}\!\left[T^l\right], \label{eq:sts4}
\end{eqnarray}
where~(\ref{eq:sts3}) is obtained by the relations~(\ref{eq:TtoS}) and~(\ref{eq:StoT}) while isolating the term $\mathbb{E}\!\left[T^{l}\right]$ and~(\ref{eq:sts4}) is established by evaluating the sum over $i$ and the integral over $r$. The proof of Lemma~\ref{lemmast} is completed by the connection between Bell polynomials and derivatives of Gamma functions as~\cite{Comtet}
\begin{eqnarray}
\left.\frac{\dd^{l-j}}{\dd a^{l-j}}\frac{\Gamma(a)}{\Gamma(mn+l)}\right\rvert_{a=mn+l}
&=&B_{l-j}\left(\psi_0(mn+l),\dots,\psi_{l-j-1}(mn+l)\right),
\end{eqnarray}
where $\psi_{k}(z)$ is the $k$-th order polygamma function~\cite{Brychkov08}
\begin{equation}\label{eq:polygamma}
\psi_{k}(z)=\frac{\dd^{k+1}}{\dd z^{k+1}}\ln\Gamma(z)=\frac{\dd^{k}}{\dd z^{k}}\psi_{0}(z)
\end{equation}
that satisfies
\begin{equation}\label{eq:psisre}
\psi_{k}(z+n)-\psi_{k}(z)=(-1)^{k}k!\sum_{i=0}^{n-1}\frac{1}{(z+i)^{k+1}}
\end{equation}
and the $k$-th complete exponential Bell polynomial $B_{k}$,
\begin{eqnarray}
B_{k}(z_1,\dots,z_k)&=&\sum_{j=1}^{k}B_{k,j}(z_1,\dots,z_{k-j+1}) \label{eq:cbp1} \\  
&=&k!\sum_{j_1+2j_2+\cdots+kj_{k}=k}\prod_{i=1}^{k}\frac{z_i^{j_i}}{(i!)^{j_i}j_{i}!}, \label{eq:cbp2}
\end{eqnarray}
is the sum of incomplete ones $B_{k,j}$.
\end{proof}

\section{Proof of Lemma~\ref{lemmart}}\label{apprt}
\begin{proof}
Inserting the Christoffel-Darboux form of one-point correlation kernel~\cite{Forrester} 
\begin{eqnarray}
K\!\left(x,x\right)&=&\frac{m!}{(m-1+\alpha)!}w(x)x^{-1}\left(mL_{m}^{(\alpha)}(x)^2+L_{m-1}^{(\alpha)}(x)L_{m}^{(\alpha)}(x)\right. \nonumber \\
&&\!\left.-(m+1)L_{m-1}^{(\alpha)}(x)L_{m+1}^{(\alpha)}(x)\right), \label{eq:kernelc3m}
\end{eqnarray}
where $L_{k}^{(\alpha)}(x)$ is the Laguerre orthogonal polynomial~\cite{Szego}
\begin{eqnarray}\label{eq:Lo}
\int_0^\infty w(x)L_l^{(\alpha)}(x)L_k^{(\alpha)}(x)\dd x=\frac{(k+\alpha)!}{k!}\delta_{lk}
\end{eqnarray}
with $w(x)$ being the corresponding weight function 
\begin{equation}\label{eq:weight}
w(x)=x^{\alpha}\ee^{-x},
\end{equation} 
into the mean value expression of $R_k$ for $k\in\mathbb{R}_{\ge0}$, cf.~(\ref{eq:Nei}),
\begin{equation}\label{eq:meanRk}
\kappa\!\left(R_k\right)=\int_0^{\infty }x^{k}K(x,x)\dd x,
\end{equation}
we have
\begin{eqnarray}\label{eq:meantkcd}
\kappa\!\left(R_k\right)&=&\frac{m!}{(m-1+\alpha)!}\Bigg(\int_{0}^{\infty}\!x^{k}w(x)L_{m-1}^{(\alpha)}(x)L_m^{(\alpha)}(x)\dd x \nonumber \\
&&+\int_0^{\infty}\!x^{k-1}w(x)\left(mL_{m}^{(\alpha)}(x)^2+(m+\alpha)L_{m-1}^{(\alpha)}(x)^2\right. \nonumber \\
&&-(2m+\alpha)\left.L_{m-1}^{(\alpha)}(x)L_{m}^{(\alpha)}(x)\right)\dd x\Bigg).
\end{eqnarray}
In arriving at the above result, we have also used the recurrence relation of Laguerre polynomials~\cite{Szego}
\begin{equation}\label{eq:recur}
xL_{m}^{(\alpha)}(x)=-(m+\alpha)L_{m-1}^{(\alpha)}(x)+(2m+1+\alpha)L_{m}^{(\alpha)}(x)-(m+1)L_{m+1}^{(\alpha)}(x).
\end{equation}

Now the task is to replace the integrals on right-hand side of~(\ref{eq:meantkcd}) to averages of $R_k$. For this purpose, we make use of the identities 
\begin{eqnarray}
\frac{m!}{(m+\alpha)!}\int_0^{\infty}x^{k}w(x)L_{m}^{(\alpha)}(x)L_{m}^{(\alpha)}(x)\dd x &=& \kappa^{\normalfont\plus}\!\left(R_k\right)-\kappa\!\left(R_k\right), \label{eq:p2meanr} \\
\frac{m!}{(m-1+\alpha)!}\int_0^{\infty}x^{k}w(x)L_{m-1}^{(\alpha)}(x)L_{m}^{(\alpha)}(x)\dd x &=& -k\kappa\!\left(R_k\right), \label{eq:p3meanr}
\end{eqnarray}
which is respectively obtained by the definitions~(\ref{eq:kernelsm}),~(\ref{eq:lm3mm}) and the integration by parts between $x^{k-1}$ and $x K\!\left(x,x\right)$ in (\ref{eq:meanRk}) when using~\cite{Forrester21}
\begin{equation}\label{eq:1pdd}
\frac{\dd}{\dd x}xK(x,x)=\frac{m!}{(m-1+\alpha)!}w(x)L_{m-1}^{(\alpha)}(x)L_m^{(\alpha)}(x).
\end{equation}
As shown in~\cite{Forrester}, the identity~(\ref{eq:1pdd}) can be established by using the recurrence relation~(\ref{eq:recur}) and the structure relation~\cite{Szego} of Laguerre polynomials 
\begin{equation}\label{eq:Lsr}
x\frac{\dd}{\dd x}L_{m}^{(\alpha)}(x)=-(m+\alpha)L_{m-1}^{(\alpha)}(x)+mL_{m}^{(\alpha)}(x).
\end{equation}
Inserting~(\ref{eq:p2meanr}) and~(\ref{eq:p3meanr}) into~(\ref{eq:meantkcd}), the recurrence relation of $\kappa\!\left(R_k\right)$ in~(\ref{eq:rrecur}) is established. Taking the derivative of~(\ref{eq:rrecur}) with respect to $k$ leads to the recurrence relation of $\kappa\!\left(T_k\right)$ in~(\ref{eq:Tkrecur}). 

The remaining task is to show the initial conditions~(\ref{eq:iR}) and~(\ref{eq:iT}), where the former~(\ref{eq:iR}) directly follows from the definition~(\ref{eq:rk}) and the latter~(\ref{eq:iT}) is derived below. Recall the definition~(\ref{eq:tk}),
\begin{equation}
T_0=\sum_{i=1}^{m}\ln x_i,
\end{equation}
the cumulant generating function of $T_0$ is
\begin{eqnarray}
K(t)&=&\ln\mathbb{E}\!\left[\ee^{tT_0}\right] \\
&=&\ln\int_{[0,\infty)^m}\frac{1}{C_\alpha}\prod_{1\leq i<j\leq m}\left(x_{i}-x_{j}\right)^{2}\prod_{i=1}^{m}x_i^{\alpha+t}\ee^{-x_i}\dd x_i \\
&=&\ln C_{\alpha+t}-\ln C_\alpha,
\end{eqnarray}
where the normalization constant $C_\alpha$ is given by $C_{\alpha,1}$ in (\ref{eq:cab}) as
\begin{equation}\label{exit_c1}
C_\alpha=\pi^{\frac{1}{2}m(m-1)}\prod_{i=0}^{m-1}\Gamma(m+\alpha-i).
\end{equation}
The $l$-th cumulant $\kappa_{l}(T_0)$ is expressed as
\begin{eqnarray}
\kappa_l(T_0)&=&\left.\frac{\dd^l}{\dd t^l}K(t)\right\rvert_{t=0} \\
&=&\sum_{k=0}^{m-1}\left.\frac{\dd^l}{\dd t^l}\ln\Gamma(m+\alpha+t-k)\right\rvert_{t=0}, \label{eq:ks2} 
\end{eqnarray}
where setting $l=1$ before applying the summation identity~\cite{Wei17}
\begin{equation}
\sum_{k=1}^{m}\psi_0(k+\alpha)=(m+\alpha)\psi_{0}(m+\alpha)-\alpha\psi_0(\alpha)-m
\end{equation}
while keeping in mind the definition~(\ref{eq:polygamma}) establishes~(\ref{eq:iT}). Finally, we note that cumulants of $T_0$ for ensembles that include the considered one~(\ref{eq:we}) as a special case have been derived in~\cite{WLLB19}. This completes the proof of Lemma~\ref{lemmart}.   
\end{proof}

\section{Proof of Proposition~\ref{propdab}} \label{appa:propdab} 
\begin{proof}
Consider a generalized Wishart density~\cite{Mathai}
\begin{equation}\label{eq:bwem}
\frac{1}{C_{\alpha,\beta}}\det\!^{\alpha}\!\left(\mathbf{ZZ}^\dagger\right)\ee^{-\beta\tr\left(\mathbf{ZZ}^\dagger\right)}~\dd\mathbf{ZZ}^\dagger
\end{equation}
of non-negative real parameters $\alpha$ and $\beta$, where the normalization constant $C_{\alpha,\beta}$ is given by
\begin{equation}\label{eq:cab}
C_{\alpha,\beta}=\frac{\pi^{\frac{1}{2}m(m-1)}}{\beta^{m(m+\alpha)}}\prod_{k=0}^{m-1}\Gamma(m+\alpha-k).
\end{equation}
For the special case $\beta=1$ and $\alpha=n-m$, the generalized density (\ref{eq:bwem}) reduces to the density of a Wishart matrix $\mathbf{ZZ}^\dagger$ with $\mathbf{Z}$ being an $m\times n$ matrix of independent complex Gaussian entries~\cite{Forrester}. In this case, the joint density of eigenvalues of the Wishart matrix is given by~(\ref{eq:we}). Introducing the generalized Wishart density~(\ref{eq:bwem}) of continuous parameters $\alpha$ and $\beta$ is necessary to find relations of derivatives of joint cumulant $\kappa_l({\bf X})$ with respect to the parameters. 

The generating function $K_l({\bf{t}})$, ${\bf t}=\{t_1,\dots,t_l\}$, of the joint cumulant $\kappa_l({\bf X})$ is defined as
\begin{equation}\label{eq:ktg}
K_l({\bf{t}})=\ln\mathbb{E}\!\left[\ee^{\sum_{i=1}^{l}t_iX_i}\right],
\end{equation}
where the cumulant is recovered as 
\begin{equation}\label{eq:ktd}
\left.\frac{\dd}{\dd t_1\cdots\dd t_l}K_l({\bf{t}})\right\rvert_{t_1=\dots=t_l=0}=\kappa_l({\bf X}).
\end{equation}
Accordingly, we denote 
\begin{equation}\label{eq:genak}
\widetilde{K}_l({\bf{t}})=\frac{\dd}{\dd\alpha}K_l({\bf{t}})
\end{equation} 
as the generating function of $\frac{\dd}{\dd\alpha}\kappa_l({\bf{X}})$, where
\begin{equation}\label{eq:ktda}
\left.\frac{\dd}{\dd t_1\cdots\dd t_l}\widetilde{K}_l({\bf{t}})\right\rvert_{t_1=\dots=t_l=0}=\frac{\dd}{\dd\alpha}\kappa_l({\bf{X}}).
\end{equation}
The parameter derivative in~(\ref{eq:genak}) is now computed as
\begin{eqnarray}
\widetilde{K}_l({\bf{t}}) &=& \frac{1}{\mathbb{E}\!\left[\ee^{\sum_{i=1}^{l}t_iX_i}\right]}\frac{\dd}{\dd\alpha}C_{\alpha,\beta}~\mathbb{E}\!\left[\ee^{\sum_{i=1}^{l}t_iX_i}\right]\frac{1}{C_{\alpha,\beta}} \\
&=& \frac{\dd}{\dd t_{l+1}}\ln\mathbb{E}\!\left.\!\left[\ee^{\sum_{i=1}^{l}t_iX_i+t_{l+1}T_0}\right]\right\rvert_{t_{l+1}=0}+C_{\alpha,\beta}~\frac{\dd}{\dd\alpha}\frac{1}{C_{\alpha,\beta}}, \label{eq:genak1}
\end{eqnarray}
where we have used the fact that
\begin{equation}\label{eq:dett}
\frac{\dd}{\dd\alpha}\det\!^{\alpha}\!\left(\mathbf{ZZ}^\dagger\right)=T_0\det\!^{\alpha}\!\left(\mathbf{ZZ}^\dagger\right)
\end{equation}
with $T_0$ defined in~(\ref{eq:tk}) being generated as
\begin{equation}
T_0=\sum_{i=1}^{m}\ln x_i=\ln\det\!\left(\mathbf{ZZ}^\dagger\right).     
\end{equation}
Inserting~(\ref{eq:genak1}) into~(\ref{eq:ktda}), one immediately establishes the claimed relation~(\ref{eq:matda}). This completes the proof of Proposition~\ref{propdab}.

In the same manner, by using the fact that 
\begin{equation}\label{eq:detr}
\frac{\dd}{\dd\beta}\ee^{-\beta\tr\left(\mathbf{ZZ}^\dagger\right)}=-R\ee^{-\beta\tr\left(\mathbf{ZZ}^\dagger\right)},
\end{equation}
one analogously arrives at 
\begin{eqnarray}
\frac{\dd}{\dd\beta}K_l({\bf{t}})=-\frac{\dd}{\dd t_{l+1}}\ln\mathbb{E}\!\left.\!\left[\ee^{\sum_{i=1}^{l}t_iX_i+t_{l+1}R}\right]\right\rvert_{t_{l+1}=0}+C_{\alpha,\beta}~\frac{\dd}{\dd\beta}\frac{1}{C_{\alpha,\beta}},
\end{eqnarray}
which leads to the result~(\ref{eq:matdb}). 

Finally, we note that the derivative relations~(\ref{eq:matda}) and~(\ref{eq:matdb}) can also be established by using the moment to cumulant relation~(\ref{eq:kumu}). The idea is to rewrite, by using the fact that
\begin{eqnarray}
\frac{\mathrm{d}}{\mathrm{d}\alpha}\mathbb{E}\!\left[X\right]&=&\mathbb{E}\!\left[XT_0\right]-\mathbb{E}\!\left[X\right]\mathbb{E}\!\left[T_0\right], \label{eq:k1ta} \\
\frac{\mathrm{d}}{\mathrm{d}\beta}\mathbb{E}\!\left[X\right]&=&-\mathbb{E}\!\left[XR\right]+\mathbb{E}\!\left[X\right]\mathbb{E}\!\left[R\right], \label{eq:k1tb}
\end{eqnarray}
the derivatives of $\kappa_{l}\!\left(\mathbf{X}\right)$ in~(\ref{eq:kumu}) with respect to $\alpha$ and $\beta$ into respective cumulants $\kappa_{l+1}\!\left(\mathbf{X},T_0\right)$ and $\kappa_{l+1}\!\left(\mathbf{X},R\right)$ of consecutive order. 
\end{proof}

\section{Proof of Lemma~\ref{lemmac}}\label{appa:lemmac}
\begin{proof}
For the joint moment of $l$ arbitrary linear statistics ${\bf X}=\left\{X_1,\ldots,X_l\right\}$,
\begin{equation}\label{eq:muxe}
\mathbb{E}\!\left[\prod_{j=1}^{l}X_j\right]=\mathbb{E}\!\left[\sum_{i_1=1}^m f_1\!\left(x_{i_1}\right)\cdots \sum_{i_l=1}^m f_l\!\left(x_{i_l}\right)\right],
\end{equation}
the expectation is written as a sum over partitions 
\begin{equation}\label{eq:mmp}
\sum_{\left\{\hspace{-0.1em}M_1,\ldots,M_i\right\}\in \mathcal{P}_{L}} \frac{m!}{(m-i)!}\mathbb{E}\!\left[\prod_{j=1}^{i}\prod_{r\in M_j}f_r\!\left(x_j\right)\right],
\end{equation}
where the set $M_j$ of each partition collects coinciding indices among $i_1,\ldots,i_l$ in~(\ref{eq:muxe}). Inserting (\ref{eq:mmp}) and the joint density $g_{N}(x_{1},\dots,x_{N})$ of $N\leq m$ arbitrary eigenvalues~\cite{Forrester} of the Wishart-Laguerre ensemble~(\ref{eq:we}),
\begin{equation}\label{eq:Nei}
g_{N}(x_{1},\dots,x_{N})=\frac{(m-N)!}{m!}\det\left(K\!\left(x_{i},x_{j}\right)\right)_{i,j=1}^{N},
\end{equation}
into (\ref{eq:muxe}), we have
\begin{equation}\label{eq:mux}
\mathbb{E}\!\left[\prod_{j=1}^{l}X_j\right]=\!\sum_{\left\{\hspace{-0.1em}M_1, \ldots, M_i\right\}\in \mathcal{P}_{L}}  \int_{[0,\infty)^i}\!\!\left(\prod_{j=1}^{i}\prod_{r\in M_j}f_r\!\left(x_j\right)\right)\det\left(K\!\left(x_{s},x_{t}\right)\right)_{s,t=1}^{i}\prod_{j=1}^{i}\dd x_j.
\end{equation}

Rewriting the permutation in the definition of determinant
\begin{equation}
\det\left(K\!\left(x_{s},x_{t}\right)\right)_{s,t=1}^{i}=\sum_{\sigma \in S_i}\text{sgn}(\sigma)\prod_{j=1}^{i}K\!\left(x_j,x_{\sigma(j)}\right)
\end{equation}
as a product of cyclic permutations~\cite{Soshnikov}, the above determinant becomes
\begin{equation}\label{eq:kcpm}
\sum_{\{\hspace{-0.1em}K_1,\dots,K_{\hspace{-0.05em}s}\}\in\mathcal{P}_I}\prod_{r=1}^s (-1)^{\rvert K_{\hspace{-0.05em}r}\rvert-1}\!\sum_{\sigma\in C_{K_{\hspace{-0.05em}r}}}\prod_{j\in K_{\hspace{-0.05em}r}} K\!\left(x_{j},x_{\sigma(j)}\right),
\end{equation}
where $\text{sgn}(\sigma)$ is sign of the permutation, $I=\{1,\dots,i\}$, and $C_{p}$ denotes the set of all cyclic permutations of set $p$. Inserting~(\ref{eq:kcpm}) into~(\ref{eq:mux}), we now change the order of partitions of $\mathcal{P}_L$ and $\mathcal{P}_{I}$ by constructing a new partition $\left\{\hspace{-0.1em}P_1,\ldots,P_s\right\}$ of $L$ as $P_q=\bigcup_{r\in K_q}\!M_r,~q=1,\ldots,s$, where $\left\{M_j:j\in K_q\right\}$ is a partition $\left\{p_1,\dots,p_{i}\right\}$ of $P_q$. Therefore, we obtain
\begin{eqnarray}\label{eq:muke}
\mathbb{E}\!\left[\prod_{j=1}^{l}X_j\right]&=&\sum_{\left\{\hspace{-0.1em}P_1,\ldots,P_s\right\}\in\mathcal{P}_L}\prod_{q=1}^s\sum_{\left\{p_1,\dots,p_{i}\right\}\in\mathcal{P}_{P_q}}\!(-1)^{i-1}\sum_{\sigma\in C_{I}}\int_{[0,\infty)^i} \prod_{j=1}^{i}\nonumber\\
&&\times\prod_{r\in p_{j} }f_r\!\left(x_j\right)K\!\left(x_{j},x_{\sigma\left(j\right)}\right)\dd x_{j}.
\end{eqnarray}

On the other hand, it is a well-known fact~\cite{PT11} that joint moments $\mathbb{E}\!\left[\prod_{j=1}^{l}X_j\right]$ can be expressed in terms of joint cumulants $\kappa_l({\bf X})$ and vice versa as
\begin{eqnarray}
\mathbb{E}\!\left[\prod_{j=1}^{l}X_j\right]&=&\sum_{\left\{\hspace{-0.1em}P_1,\ldots,P_s\right\}\in\mathcal{P}_L}\prod_{q=1}^{s}\kappa_{\rvert P_q\rvert}\!\left({\bf X}_{P_q}\right), \label{eq:muku}\\
\kappa_l({\bf X})&=&\sum_{\left\{\hspace{-0.1em}P_1,\ldots, P_s\right\}\in\mathcal{P}_L}(-1)^{s-1}(s-1)!\prod_{q=1}^{s}\mathbb{E}\!\left[\prod_{j\in P_q} X_j\right]. \label{eq:kumu}
\end{eqnarray}
For the special case of identical statistics $X=X_1=\cdots=X_l$ with the shorthand notations $\mu_l=\mathbb{E}\!\left[X^l\right]$ and $\kappa_l=\kappa_l(X)$, one has
\begin{eqnarray}
\mu_l&=&\sum_{k=1}^{l}B_{l,k}\left(\kappa_1,\ldots,\kappa_{l-k+1}\right), \label{eq:mk} \\
\kappa_l&=&\sum_{k=1}^l(-1)^{k-1}(k-1)!B_{l, k}\left(\mu_1, \ldots, \mu_{l-k+1}\right), \label{eq:km}
\end{eqnarray}
where $B_{l,k}$ is the incomplete Bell polynomials~(\ref{eq:cbp1}). Now comparing~(\ref{eq:muke}) with~(\ref{eq:muku}) for the case $s=1$, the joint cumulant $\kappa_l\!\left({\bf X}\right)$ is expressed as
\begin{eqnarray}\label{eq:jcuker1}
\!\!\!\!\!\!\!\kappa_l\!\left({\bf X}\right)=\!\!\sum_{\left\{p_{1}, \ldots, p_{i}\right\}\in \mathcal{P}_L}(-1)^{i-1}\!\sum_{\sigma\in C_{I}}\int_{[0,\infty)^{i}}\prod_{j=1}^{i}\prod_{r\in p_{j}}f_r\!\left(x_j\right)~\!\!K\!\left(x_j,x_{\sigma(j)}\right)\dd x_j.      
\end{eqnarray}

The final piece of the proof is to replace in~(\ref{eq:jcuker1}) the summation of kernels over cyclic permutations by summation of kernels over permutations 
\begin{equation}\label{eq:kctp}
\sum_{\sigma\in C_I}\prod_{j=1}^{i}K\!\left(x_j,x_{\sigma(j)}\right)=\frac{1}{i}\sum_{\sigma\in S_i}\prod_{j=1}^{i}K\!\left(x_{\sigma(j)},x_{\sigma(j+1)}\right).
\end{equation}
Lemma~\ref{lemmac} is then proved when applying the inverse permutation of $\sigma$ that corresponds to relabeling of indices. 
\end{proof}

\section{Proof of Proposition~\ref{props}}\label{appa:props}
\begin{proof} 
Inserting into~(\ref{eq:jcuker1}) the summation form of correlation kernel~(\ref{eq:ksum}),
\begin{equation}\label{eq:kernelsm}
K\!\left(x,y\right)=\sqrt{w(x)w(y)}\sum_{k=0}^{m-1}\frac{k!}{(k+\alpha)!}L_k^{(\alpha)}(x)L_k^{(\alpha)}(y),
\end{equation} 
with $w(x)$ being the weight function~(\ref{eq:weight}) and $L_k^{(\alpha)}(x)$ being the Laguerre polynomial~(\ref{eq:Lo}), it follows that
\begin{eqnarray}\label{eq:p21}
&&\kappa_l^{\plus}({\bf X})-\kappa_l({\bf X})
=\sum_{\left\{\hspace{-0.1em}M_1,\ldots,M_i\right\}\in\mathcal{P}_{L}}(-1)^{i-1}\sum_{\sigma\in C_I }\int_{[0,\infty)^{i}}\prod_{j=1}^{i}\prod_{r\in M_{j}}f_r\!\left(x_j\right) \nonumber \\
&&\times\!\left(\prod_{j=1}^{i}\left(K\!\left(x_j,x_{\sigma(j)}\right)+ L_{\hspace{-0.1em}H}\!\left(x_j,x_{\sigma(j)}\right)\right)-\prod_{j=1}^{i}K\!\left(x_j,x_{\sigma(j)}\right)\!\right)\prod_{j=1}^{i}\dd x_j,
\end{eqnarray}
where 
\begin{equation}
L_{\hspace{-0.1em}H}\!\left(x,y\right)=\frac{m!}{(m+\alpha)!}\sqrt{w(x)w(y)}L_m^{(\alpha)}(x)L_m^{(\alpha)}(y)
\end{equation}
as also defined in~(\ref{eq:LH}).

Let $s$ be the number of times the term $L_{\hspace{-0.1em}H}\!\left(x_j,x_{\sigma(j)}\right)$ is picked up from the product of kernels in~(\ref{eq:p21}),
\begin{equation}
\prod_{j=1}^{i}\left(K\!\left(x_j,x_{\sigma(j)}\right)+ L_{\hspace{-0.1em}H}\!\left(x_j,x_{\sigma(j)}\right)\right),
\end{equation}
each of which is interpreted as a partition of set $\{1,\ldots,i\}$ into $s$ subsets. This interpretation is based on the fact that the presence of products of $L_{\hspace{-0.1em}H}\!\left(x,y\right)$ decouples the integral in~(\ref{eq:p21}) into products of integrals in the same manner as a partition of the set. Consequently, the summation over cyclic permutations in~(\ref{eq:p21}) becomes
\begin{eqnarray}
\!\!\!\!\!\!\!\!\!\!\!\!\!\!&&\sum_{\sigma \in C_I}\!\!\left(\prod_{j=1}^{i}\left(K\!\left(x_j,x_{\sigma(j)}\right)+ L_{\hspace{-0.1em}H}\!\left(x_j,x_{\sigma(j)}\right)\right)-\prod_{j=1}^{i}K\!\left(x_j,x_{\sigma(j)}\right)\!\right) \nonumber \\
\!\!\!\!\!\!\!\!\!\!\!\!\!\!&=&\sum_{\{\hspace{-0.1em}K_1,\dots,K_{\hspace{-0.05em}s}\} \in \mathcal{P}_I}\!\!(s-1)!\prod_{\lambda=1}^{s}\sum_{\sigma\in C_{K_{\hspace{-0.05em}\lambda}}}\sum_{r\in K_{\hspace{-0.05em}\lambda}} L_{\hspace{-0.1em}H}\!\left(x_{r},x_{\sigma(r)}\right)\!\!\prod_{j\in K_{\hspace{-0.1em}\lambda}\setminus\{r\}}\!\!K\!\left(x_{j}, x_{\sigma(j)}\right). \label{eq:p22}
\end{eqnarray}
Inserting (\ref{eq:p22}) into (\ref{eq:p21}) before using the fact that
\begin{equation}
(-1)^{i-1}=(-1)^{s-1}\prod_{\lambda=1}^{s}(-1)^{\rvert K_{\hspace{-0.05em}\lambda}\rvert-1},
\end{equation}
the resulting expression now allows the change of the order of partitions between $\left\{\hspace{-0.1em}M_1, \ldots, M_i\right\}$ and $\left\{\hspace{-0.1em}K_1,\dots,K_{\hspace{-0.05em}s}\right\}$ as similarly performed in~(\ref{eq:mux})--(\ref{eq:muke}). We have
\begin{eqnarray}
\kappa_l^{\plus}({\bf X})-\kappa_l({\bf X})=
\sum_{\left\{\hspace{-0.1em}P_1,\ldots,P_s\right\}\in\mathcal{P}_L}(-1)^{s-1}(s-1)!\prod_{q=1}^s H_{\rvert P_q\rvert}\!\left({\bf X}_{P_q}\right),\label{eq:csH}
\end{eqnarray}
where 
\begin{eqnarray}
H_{\rvert P_q\rvert}\!\left({\bf X}_{P_q}\right)&=&\sum_{\left\{p_1,\ldots,p_i\right\}\in\mathcal{P}_{P_q}}(-1)^{i-1}\sum_{\sigma\in C_{P_q}}\sum_{r=1}^{i}\int_{[0,\infty)^i}\!\!L_H\!\left(x_r,x_{\sigma(r)}\right) \nonumber \\
&&\times\prod_{j=1}^{i}\prod_{s\in p_{j}}f_s\!\left(x_j\right)\prod_{1\leq j\neq r\leq i} K\!\left(x_j, x_{\sigma(j)}\right)\prod_{j=1}^{i}\dd x_{j}\label{eq:exitdic}
\end{eqnarray}
is in fact the decoupled integral defined in~(\ref{eq:dcIth}). Finally, comparing~(\ref{eq:csH}) with the moments to cumulants relation~(\ref{eq:kumu}), the claimed result~(\ref{eq:Hcs}) is established by the dual relation~(\ref{eq:muku}) from cumulants to moments.

The result~(\ref{eq:hcs}) is shown analogously. In this case, the cumulant difference $\kappa_l^{\minus}({\bf X})-\kappa_l({\bf X})$ can be written in the same form as (\ref{eq:p21}) with $L_{\hspace{-0.1em}H}\!\left(x_j,x_{\sigma(j)}\right)$ replaced by $-L_h\!\left(x_j,x_{\sigma(j)}\right)$ defined in~(\ref{eq:Lh}). This completes the proof of Proposition~\ref{props}.
\end{proof}

\section{Proof of Proposition~\ref{propdk}}\label{appa:propdk}
\begin{proof}
The idea of the proof is to construct appropriate integration by parts when applying the operator
\begin{equation}\label{eq:op}
B_k=\sum_{s=1}^{k}B_{x_s,x_{s+1}},
\end{equation}
where $x_{k+1}=x_{1}$ and 
\begin{equation}
B_{x,y}=1+x\frac{\dd}{\dd x}+y\frac{\dd}{\dd y},
\end{equation} 
on correlation kernels. For the Laguerre kernel $K(x,y)$ in~(\ref{eq:Nei}), it is a known result that~\cite{Forrester,Forrester21}
\begin{equation}\label{eq:2pd}
B_{x,y}K(x,y)=-L_{\hspace{-0.1em}D}\!\left(x,y\right),
\end{equation}
where 
\begin{equation}
L_{\hspace{-0.1em}D}\!\left(x,y\right)=-\frac{m!\sqrt{w(x)w(y)}}{2(m-1+\alpha)!}\left(L_{m}^{(\alpha)}(x)L_{m-1}^{(\alpha)}(y)+L_{m-1}^{(\alpha)}(x) L_{m}^{(\alpha)}(y)\right)
\end{equation}
as defined in~(\ref{eq:LD}). Consequently, one has
\begin{eqnarray}
&&B_i\prod_{j=1}^{i} K\!\left(x_j,x_{j+1}\right) \nonumber \\
&=&-2\sum_{r=1}^{i}L_{\hspace{-0.1em}D}\!\left(x_r,x_{r+1}\right)\!\prod_{1\leq j\neq r\leq i}K\!\left(x_j,x_{j+1}\right)-i\prod_{j=1}^{i} K\!\left(x_j,x_{j+1}\right). \label{eq:d2p3}
\end{eqnarray}

The combinatorial structure of the decoupled integral $D_l\!\left({\bf{X}}\right)$ in~(\ref{eq:dcIth}) can be also written in terms of sum over permutations by using~(\ref{eq:kctp}) as
\begin{eqnarray}\label{eq:exitdi}
D_l\!\left({\bf{X}}\right)
&=&\sum_{\{p_1,\dots,p_i\}\in\mathcal{P}_{L}}\sum_{\sigma\in S_{i}}\frac{(-1)^{i-1}}{i} \nonumber \\
&&\times\int_{[0,\infty)^i}\sum_{r=1}^{i}L_{\hspace{-0.1em}D}\!\left(x_r,x_{r+1}\right)\prod_{1\leq j \neq r \leq i} \!K\!\left(x_j, x_{j+1}\right)\prod_{j=1}^{i}F_{j}\dd x_j,
\end{eqnarray}
where $F_j=\prod_{r\in p_j}f_r\!\left(x_{\sigma\left(j\right)}\right)$. Inserting the right-hand side of (\ref{eq:d2p3}) into (\ref{eq:exitdi}), we have
\begin{eqnarray}\label{eq:prop1ipl1}
-2D_l\!\left({\bf{X}}\right)
&=&\sum_{\left\{p_1,\ldots,p_i\right\}\in\mathcal{P}_{L}}\sum_{\sigma\in S_{i}}\frac{(-1)^{i-1}}{i}\int_{[0,\infty)^i}\prod_{j=1}^{i}F_jB_i\prod_{j=1}^{i}K\!\left(x_j,x_{j+1}\right)\dd x_j \nonumber \\
&&+\sum_{\left\{p_1,\ldots,p_i\right\}\in\mathcal{P}_{L}}\sum_{\sigma\in S_{i}}(-1)^{i-1}\int_{[0,\infty)^i}\prod_{j=1}^{i}F_{j}K\!\left(x_j,x_{j+1}\right)\dd x_j.
\end{eqnarray}

We now perform integration by parts on the first integral of the right-hand side of~(\ref{eq:prop1ipl1}). For any functions $f$ and $g$ of $i$ variables, one has the skew self-adjoint property of the operator $B_{x,y}$, hence that of $B_{i}$, as~\cite{Forrester,Forrester21}
\begin{equation}\label{eq:sksm}
\int f B_{i} g~\!\prod_{j=1}^{i}\dd x_j = -\int g B_{i} f~\!\prod_{j=1}^{i} \dd x_j
\end{equation}
whenever the integrals in~(\ref{eq:sksm}) exist and the product $fg$ vanishes on the boundary of support. Applying~(\ref{eq:sksm}) in (\ref{eq:prop1ipl1}) with 
\begin{eqnarray}
f &=& \prod_{j=1}^{i}F_j, \\
g &=& \prod_{j=1}^{i}K\!\left(x_j,x_{j+1}\right)
\end{eqnarray}
while keeping in mind the joint cumulant expression~(\ref{eq:jcuker}), the right-hand side of~(\ref{eq:prop1ipl1}) becomes $-2\kappa_l^{\prime}({\bf X})$. This completes the proof of Proposition~\ref{propdk}.
\end{proof}

\section{Proof of Corollary~\ref{codkb}}\label{appa:coro1}
\begin{proof}
Besides the result~(\ref{eq:redk}), the proof of Corollary~\ref{codkb} also relies on the result
\begin{equation}\label{eq:dbR}
\kappa_{l+1}({\bf X},R)=\kappa_{l}^{\prime}({\bf X}).
\end{equation}
To show~(\ref{eq:dbR}), it is convenient to consider the generalized Wishart density~(\ref{eq:bwem}), the corresponding correlation kernel $K_{\beta}(x,y)$ is expressed through that of the Wishart density as
\begin{equation}\label{eq:kernelsmb}
K_{\beta}(x,y)=\beta K(\beta x,\beta y).
\end{equation}
The fact that
\begin{equation}
\beta\frac{\dd}{\dd\beta}K(\beta x,\beta y)=\left(x\frac{\dd}{\dd x}+y\frac{\dd}{\dd y}\right)K(\beta x,\beta y)
\end{equation}
leads to
\begin{equation}\label{eq:p32b}
\beta\frac{\dd}{\dd \beta}\frac{1}{\beta^i}\prod_{j=1}^{i} K_{\beta}\!\left(x_j,x_{j+1}\right)=\frac{1}{\beta^i}\left(\sum_{j=1}^{i}x_j\frac{\dd}{\dd x_j}\right)\prod_{j=1}^{i} K_{\beta}\!\left(x_j,x_{j+1}\right),
\end{equation}
where by expressing the right-hand side of the above in terms of the operator $B_i$ in (\ref{eq:op}) before setting $\beta=1$, we have
\begin{equation}\label{eq:dbdx0}
B_i\prod_{j=1}^{i} K(x_j,x_{j+1})
=2\frac{\dd}{\dd \beta}\prod_{j=1}^{i} K_{\beta}\!\left(x_j,x_{j+1}\right)\Big\rvert_{\beta=1}-i\prod_{j=1}^{i} K(x_j,x_{j+1}).
\end{equation}
Inserting~(\ref{eq:dbdx0}) into~(\ref{eq:prop1ipl1}) and comparing the resulting integral with the joint cumulant expression $\kappa_{l}({\bf X})$ in Lemma~\ref{lemmac}, one has
\begin{equation}\label{eq:redb}
-D_l\!\left({\bf{X}}\right)=\frac{\dd}{\dd\beta}\kappa_{l}({\bf X})\rvert_{\beta=1}.
\end{equation}
Combining the above result~(\ref{eq:redb}) with (\ref{eq:redk}) leads to
\begin{equation}\label{eq:matdbk}
\frac{\dd}{\dd\beta}\kappa_l({\bf X})\rvert_{\beta=1}=-\kappa^\prime_l({\bf X}),
\end{equation}
which, after applying the derivative relation~(\ref{eq:matdb}), establishes the claimed result~(\ref{eq:dbR}).

The rest of the proof proceeds as a direct consequence of the results~(\ref{eq:redk}) and~(\ref{eq:dbR}). By the definitions~(\ref{eq:tk}), (\ref{eq:rk}), and~(\ref{eq:xprime}), one first notices that 
\begin{eqnarray}
T_k^\prime&=&k T_k+R_k, \\
R_k^\prime&=&k R_k.
\end{eqnarray}
The result~(\ref{eq:redk}) now yields
\begin{eqnarray}
D_l(T_k,T,\dots,T)&=&\kappa_l^{\prime}(T_k,T,\dots,T) \label{eq:corokb1} \\
&=&(k+l-1)\kappa_l(T_k,T,\dots,T)+\kappa_l(R_k,T,\dots,T) \nonumber \\
&&+~\!(l-1)\kappa_{l}(T_k,T,\dots,T, R), \label{eq:corokb2}
\end{eqnarray}
where the last term is then written by using~(\ref{eq:dbR}) as  
\begin{equation}\label{eq:corokb3}
\kappa_{l}(T_k,T,\dots,T,R)=\kappa_{l-1}^\prime(T_k,T,\dots,T).
\end{equation}
Continuing to iterate $l-1$ more times the procedure from~(\ref{eq:corokb1}) to~(\ref{eq:corokb3}) leads to the claimed result~(\ref{eq:coro11}). The decoupled integral $D_l(R_k,T,\dots,T)$ is processed analogously. This completes the proof of Corollary~\ref{codkb}. 
\end{proof}

\section{Proof of Theorem~\ref{Theorem1}}\label{appa:Theorem1}
\begin{proof}
The starting point of the proof is to show, by using Proposition~\ref{propdab}, that the decoupling structure~(\ref{eq:th1}) ensures decoupling of correlation kernels in a summation-free manner by producing necessary factors to cancel the denominators of a pair of Christoffel-Darboux kernels~(\ref{eq:kernelcm}). Subsequently, it is shown that the resulting decoupled term $\delta_l(k)$ can be written in terms of lower-order cumulants~(\ref{eq:dT}) by making use of Propositions~\ref{props},~\ref{propdk} and Corollary~\ref{codkb}.

Using Proposition~\ref{propdab} that gives rise to the decoupling statistics 
\begin{equation}\label{eq:destat}
\frac{\dd}{\dd\alpha}\kappa_{l-1}(T_{k+1},T,\dots,T)=\kappa_{l}(T_{k+1},T,\dots,T,T_0),    
\end{equation}
the decoupled term is 
\begin{eqnarray}
\delta_l(k)&=&\kappa_{l}(T_k,T,\dots,T)-\frac{\dd}{\dd\alpha}\kappa_{l-1}(T_{k+1},T,\dots,T)\\
&=&\frac{1}{2}\big(\kappa_{l}(T_k,T,\dots,T)-\kappa_{l}(T_{k+1},T,\dots,T,T_0)\nonumber\\
&&+~\!\kappa_{l}(T,T,\dots,T_k)-\kappa_{l}(T_{0},T,\dots,T,T_{k+1})\big), \label{eq:dlk4}
\end{eqnarray}
where it is shown below that~(\ref{eq:dlk4}) permits the decoupling of kernels by factoring desired terms from the differences of joint cumulants in~(\ref{eq:dlk4}). Inserting the cumulant expression $\kappa_l({\bf X})$ in~(\ref{eq:jcuker1}) into (\ref{eq:dlk4}) and by using the fact that
\begin{equation}
x_{t_1}^k-x_{t_2}^k=\sum_{r=0}^{k-1}x_{t_1}^rx_{t_2}^{k-r-1}(x_{t_1}-x_{t_2}),
\end{equation}
one has 
\begin{eqnarray}\label{eq:dckp01} 
\delta_l(k)&=&\frac{1}{2}\sum_{r=0}^{k-1}\sum_{\left\{\hspace{-0.1em}M_1,\ldots,M_i\right\}\in\mathcal{P}_L}(-1)^{i-1}\int_{[0,\infty)^i}\left(x_{t_1}- x_{t_2}\right)\left(x_{t_2}-x_{t_1}\right)\nonumber\\
&&\times\sum_{\sigma\in C_I}\prod_{j=1}^{i}\prod_{s\in M_j}f_s\!\left(x_{j}\right)K\!\left(x_j,x_{\sigma(j)}\right)\dd x_j,
\end{eqnarray}
where 
\begin{eqnarray}
f_1\!\left(x\right) &=& x^{r}\ln x,\label{eq:dckp01f1} \\
f_{2}(x) &=& \cdots=f_{l-1}(x)=x\ln x, \\
f_l\!\left(x\right) &=& x^{k-r-1}\ln x \label{eq:dckp01fl}
\end{eqnarray}
and the indices $t_{1}, t_{2}\in L=\{1,\dots,l\}$ with $M_{t_1}$ and $M_{t_2}$ denoting respectively subsets of a partition of $L$ containing $1$ and $l$.

To see the decoupling of kernels in~(\ref{eq:dckp01}), we first write the differences $x_{t_1}-x_{t_2}$ and $x_{t_2}-x_{t_1}$ of a cyclic permutation $\sigma\in C_I$ in~(\ref{eq:dckp01}) as sum of differences $x_j-x_{\sigma(j)}$,
\begin{eqnarray}
&&x_{t_1}-x_{t_2}=x_{t_1}-x_{\sigma(t_1)}+x_{\sigma(t_1)}-x_{\sigma(\sigma(t_1))}+\dots+x_{\sigma^{-1}(t_2)}-x_{t_2}, \\
&&x_{t_2}-x_{t_1}=x_{t_2}-x_{\sigma(t_2)}+x_{\sigma(t_2)}-x_{\sigma(\sigma(t_2))}+\dots+x_{\sigma^{-1}(t_1)}-x_{t_1}.
\end{eqnarray}
A pair of kernels in~(\ref{eq:dckp01}) are then decoupled by using the Christoffel-Darboux form~(\ref{eq:kernelcm}) to cancel the product of factors $\left(x_{j_1}-x_{\sigma\left(j_1\right)}\right)\left(x_{j_2}-x_{\sigma\left(j_2\right)}\right)$ in the denominators as
\begin{eqnarray}
&&\left(x_{j_1}-x_{\sigma\left(j_1\right)}\right)\left(x_{j_2}-x_{\sigma\left(j_2\right)}\right)\prod_{j=1}^{i}\!K\!\left(x_j,x_{\sigma(j)}\right) \\
&=&\mathrm{pL}\prod_{s_1\in K_1\setminus\{j_2\}}\!\!\!\!\!\!\!\!K\!\left(x_{s_1},x_{\sigma(s_1)}\right)\prod_{s_2\in K_2\setminus\{j_1\}}\!\! K\!\left(x_{s_2},x_{\sigma(s_2)}\right) \label{eq:th1dc},
\end{eqnarray}
where $K_1$ and $K_2$ denote respectively the set of elements in cycles from $\sigma(j_1)$ to $j_2$ and $\sigma(j_2)$ to $j_1$ with $K_1\cup K_2=\{1,\dots,i\}$, and pL are the decoupled terms consisting of products of Laguerre polynomials from the numerators of Christoffel-Darboux kernels, cf. (\ref{eq:LH})--(\ref{eq:LD}). 

Having seen that $\delta_l(k)$ in (\ref{eq:th1}) permits decoupling of correlation kernels in a summation-free manner, we now show that the decoupled term $\delta_l(k)$ can be recycled into lower-order cumulants in the form of~(\ref{eq:dT}). To facilitate the change of the order of partitions, it is convenient to first divide $\delta_{l}(k)$ in~(\ref{eq:dckp01}) into three parts 
\begin{equation}\label{eq:t1p1a123}
\delta_l(k)=\frac{1}{2}\sum_{r=0}^{k-1}(A_1+A_2+A_3),   
\end{equation}
where $A_1$, $A_2$, and $A_3$ denote respectively the following three cases:
\begin{enumerate}
{\small
\item Neither $M_{t_1}$ nor $M_{t_2}$ is a singleton: $\rvert M_{t_1}\rvert>1, \rvert M_{t_2}\rvert>1$.
\item One of $M_{t_1}$ and $M_{t_2}$ is a singleton: $\rvert M_{t_1}\rvert=1, \rvert M_{t_2}\rvert>1$ or $\rvert M_{t_1}\rvert>1, \rvert M_{t_2}\rvert=1$.
\item Both $M_{t_1}$ and  $M_{t_2}$ are singletons: $\rvert M_{t_1}\rvert=1, \rvert M_{t_2}\rvert=1$.}
\end{enumerate}

By definition~(\ref{eq:dckp01}), $A_1$ is given by
\begin{eqnarray}\label{eq:A1}
A_1&=&\sum_{\left\{\hspace{-0.1em}M_1, \ldots, M_i\right\}\in \mathcal{P}_{L\setminus\{1,l\}}} (-1)^{i-1} \int_{[0,\infty)^i}\prod_{j=1}^{i}\prod_{s\in M_j}f_s\!\left(x_{j}\right)\nonumber\\
&& \times  \sum_{\sigma \in C_I }\sum_{t_1,t_2=1}^{i} \left( x_{t_1}-x_{t_2}\right)\left( x_{t_2}-x_{t_1}\right)\prod_{j=1}^{i} K\!\left(x_j,x_{\sigma(j)}\right)\dd x_j.
\end{eqnarray}
Inserting~(\ref{eq:th1dc}) into~(\ref{eq:A1}) and changing the summation over the cyclic permutations $\sigma\in C_I$ to summations over partitions $\{K_1,K_2\}\in\mathcal{P}_I$, we have
\begin{eqnarray}\label{eq:th1dcA1}
&&\sum_{\sigma \in C_I }\sum_{t_1,t_2=1}^{i} f_1(x_{t_1})f_2(x_{t_2})\left( x_{t_1}-x_{t_2}\right)\left( x_{t_2}-x_{t_1}\right)\prod_{j=1}^{i} K\!\left(x_j,x_{\sigma(j)}\right) \nonumber \\
&=&\sum_{\left\{\hspace{-0.1em}K_{1},K_{2}\right\}\in\mathcal{P}_I}\sum_{\sigma\in S_2}\sum_{t_1,s_1\in K_{\sigma(1)}}\sum_{t_2,s_2\in K_{\sigma(2)}}\sum_{\sigma_1 \in C_{K_{\sigma(1)}}}\sum_{\sigma_2 \in C_{K_{\sigma(2)}}}f_1(x_{t_1})f_2(x_{t_2}) \nonumber \\
&&\times~\!\bigg(\!-\kappa(R)L_{\hspace{-0.1em}H}\!\left(x_{s_1},x_{\sigma_1(s_1)}\right)L_h\!\left(x_{s_2},x_{\sigma_2(s_2)}\right)-\kappa(R)L_h\!\left(x_{s_1},x_{\sigma_1(s_1)}\right) \nonumber \\
&&\times~\!L_{\hspace{-0.1em}H}\!\left(x_{s_2},x_{\sigma_2(s_2)}\right)+2L_{\hspace{-0.1em}D}\!\left(x_{s_1},x_{\sigma_1(s_1)}\right)L_{\hspace{-0.1em}D}\!\left(x_{s_2},x_{\sigma_2(s_2)}\right)\!\bigg) \nonumber \\
&&\times\prod_{j_1\in K_{\sigma(1)}\setminus\{s_1\}}\!K\!\left(x_{j_1},x_{\sigma_1(j_1)}\right)\prod_{j_2\in K_{\sigma(2)}\setminus\{s_2\}} \!K\!\left(x_{j_2},x_{\sigma_2(j_2)}\right),
\end{eqnarray}
where the sum over $\sigma: \sigma(1)=1, \sigma(2)=2$ and $\sigma(1)=2, \sigma(2)=1$ of the symmetric group $S_2$ introduces the order of subsets $K_1$, $K_2$ and
\begin{eqnarray}
L_{\hspace{-0.1em}H}\!\left(x,y\right)&=&\frac{m!}{(m+\alpha)!}\sqrt{w(x)w(y)}L_m^{(\alpha)}(x)L_m^{(\alpha)}(y),\label{eq:LH}\\ 
L_{\hspace{-0.1em}h}\!\left(x,y\right)&=&\frac{(m-1)!}{(m-1+\alpha)!}\sqrt{w(x)w(y)}L_{m-1}^{(\alpha)}(x) L_{m-1}^{(\alpha)}(y),\label{eq:Lh}\\
L_{\hspace{-0.1em}D}\!\left(x,y\right)&=&-\frac{m!\sqrt{w(x)w(y)}}{2(m-1+\alpha)!}\left(L_{m}^{(\alpha)}(x) L_{m-1}^{(\alpha)}(y)+L_{m-1}^{(\alpha)}(x) L_{m}^{(\alpha)}(y)\right) \label{eq:LD}
\end{eqnarray}
denote three types of decoupled terms resulting from the decoupling of kernels with the replacement
\begin{equation}
\kappa(R)=m(m+\alpha)
\end{equation}
in~(\ref{eq:th1dcA1}) for a more compact result. The representation $(\ref{eq:th1dcA1})$ now allows the change of partition orders, similarly to the steps from~(\ref{eq:mux}) to~(\ref{eq:muke}), as
\begin{eqnarray}\label{eq:t1p1a1s}
A_1&=&\sum_{\sigma\in S_2}\sum_{\left\{\hspace{-0.1em}P_1, P_2\right\}\in \mathcal{P}_{L\setminus\{1,l\}}}\!\bigg(\!\left(H^{(1)}_{\rvert P_{\sigma(1)}\rvert+1}\!\left(X_1,{\bf X}_{P_{\sigma(1)}}\right) h^{(1)}_{\rvert P_{\sigma(2)}\rvert+1}\!\left(X_l,{\bf X}_{P_{\sigma(2)}}\right)\right.\nonumber\\
&&+~\!\left.\!\!h^{(1)}_{\rvert P_{\sigma(1)}\rvert+1}\!\left(X_1,{\bf X}_{P_{\sigma(1)}}\right) H^{(1)}_{\rvert P_{\sigma(2)}\rvert+1}\!\left(X_l,{\bf X}_{P_{\sigma(2)}}\right)\right)\kappa(R)\nonumber\\
&&-~\!2D^{(1)}_{\rvert P_{\sigma(1)}\rvert+1}\!\left(X_1,{\bf X}_{P_{\sigma(1)}}\right)D^{(1)}_{\rvert P_{\sigma(2)}\rvert+1}\!\left(X_l,{\bf X}_{P_{\sigma(2)}}\right)\!\bigg),
\end{eqnarray}
where $D_{\rvert P_q\rvert+1}^{(1)}\!\left(X_1,{\bf X}_{P_q}\right)$ denotes 
\begin{eqnarray}\label{eq:dcid}
&&D_{\rvert P_q\rvert+1}^{(1)}\!\left(X_1,{\bf X}_{P_q}\right)=\sum_{\left\{\hspace{-0.1em}p_1,\ldots,p_i\right\}\in\mathcal{P}_{P_q}}(-1)^{i-1}\sum_{\sigma\in C_{P_q}}\sum_{v=1}^{i}\int_{[0,\infty)^i}\sum_{t=1}^{i}f_1(x_t) \nonumber \\
&&\times~\!L_{\hspace{-0.1em}D}\!\left(x_v,x_{\sigma(v)}\right)\prod_{j=1}^{i}\prod_{s\in p_j}f_{s}(x_j)\prod_{1\leq j \neq v \leq i}K\!\left(x_j, x_{\sigma(j)}\right)\prod_{j=1}^{i}\dd x_j,
\end{eqnarray}
and the shorthand notations $H_{\rvert P_q\rvert+1}^{(1)}\!\left(X_1,{\bf X}_{P_q}\right)$ and $h_{\rvert P_q\rvert+1}^{(1)}\!\left(X_1,{\bf X}_{P_q}\right)$ are respectively the expression~(\ref{eq:dcid}) with $L_{\hspace{-0.1em}D}\!\left(x_r,x_{\sigma(r)}\right)$ replaced by $L_{\hspace{-0.1em}H}\!\left(x_r,x_{\sigma(r)}\right)$ and $L_{\hspace{-0.1em}h}\!\left(x_r,x_{\sigma(r)}\right)$. We remind that the functions~(\ref{eq:dckp01f1})--(\ref{eq:dckp01fl}) correspond to the linear statistics in~(\ref{eq:t1p1a1s}) as
\begin{eqnarray}
X_1&=&T_r, \label{eq:th1f1}\\
X_2&=&\dots=X_{l-1}=T, \label{eq:th1ft}\\
X_l&=&T_{k-r-1}. \label{eq:th1fl}
\end{eqnarray}

For the other two cases in~(\ref{eq:t1p1a123}), we similarly obtain
\begin{eqnarray}
A_2&=&2\sum_{\sigma\in S_2}\sum_{\left\{\hspace{-0.1em}P_1, P_2\right\}\in\mathcal{B}}\!\bigg(\!\left(H^{(1)}_{\rvert P_{\sigma(1)}\rvert+1}\!\left(X_1,{\bf X}_{P_{\sigma(1)}}\right)h^{(2)}_{\rvert P_{\sigma(2)}\rvert+1}\!\left(X_l,{\bf X}_{P_{\sigma(2)}}\right)\right.\nonumber\\
&&+~\!\left.\!\!h^{(1)}_{\rvert P_{\sigma(1)}\rvert+1}\!\left(X_1,{\bf X}_{P_{\sigma(1)}}\right)H^{(2)}_{\rvert P_{\sigma(2)}\rvert+1}\!\left(X_l,{\bf X}_{P_{\sigma(2)}}\right)\right)\kappa(R)\nonumber\\
&&-~\!2D^{(1)}_{\rvert P_{\sigma(1)}\rvert+1}\!\left(X_1,{\bf X}_{P_{\sigma(1)}}\right)D^{(2)}_{\rvert P_{\sigma(2)}\rvert+1}\!\left(X_l,{\bf X}_{P_{\sigma(2)}}\right)\!\bigg),\label{eq:t1p1a2s}\\
A_3&=&\sum_{\sigma\in S_2}\sum_{\left\{\hspace{-0.1em}P_1, P_2\right\}\in \mathcal{B}}\!\bigg(\!\left(H^{(2)}_{\rvert P_{\sigma(1)}\rvert+1}\!\left(X_1,{\bf X}_{P_{\sigma(1)}}\right)h^{(2)}_{\rvert P_{\sigma(2)}\rvert+1}\!\left(X_l,{\bf X}_{P_{\sigma(2)}}\right)\right.\nonumber\\
&&+~\!\left.\!\!h^{(2)}_{\rvert P_{\sigma(1)}\rvert+1}\!\left(X_1,{\bf X}_{P_{\sigma(1)}}\right)H^{(2)}_{\rvert P_{\sigma(2)}\rvert+1}\!\left(X_l,{\bf X}_{P_{\sigma(2)}}\right)\right)\kappa(R)\nonumber\\
&&-~\!2D^{(2)}_{\rvert P_{\sigma(1)}\rvert+1}\!\left(X_1,{\bf X}_{P_{\sigma(1)}}\right)D^{(2)}_{\rvert P_{\sigma(2)}\rvert+1}\!\left(X_l,{\bf X}_{P_{\sigma(2)}}\right)\!\bigg),\label{eq:t1p1a3s}
\end{eqnarray}
where $\mathcal{B}=\mathcal{P}_{L\setminus\{1,l\}}\union\{\varnothing,\{2,\dots,l\}\}$ with $P_1$ potentially being an empty set to accommodate cases of the decoupling in~(\ref{eq:th1dc}) when $K_1=\{t_2\}$, $\rvert M_{t_2}\rvert=1$ and $K_2=\{t_1\}$, $\rvert M_{t_1}\rvert=1$. In~(\ref{eq:t1p1a2s})--(\ref{eq:t1p1a3s}), $D_{\rvert P_q\rvert+1}^{(2)}\!\left(X_1,{\bf X}_{P_q}\right)$ denotes the decoupled term
\begin{eqnarray}\label{eq:dcid2}
&&D_{\rvert P_q\rvert+1}^{(2)}\!\left(X_1,{\bf X}_{P_q}\right)=\sum_{\left\{\hspace{-0.1em}p_1, \ldots, p_i\right\} \in \mathcal{P}_{P_q}} (-1)^{i}\sum_{\sigma \in C_{P_q\union\{1\}}}\sum_{v=1}^{i+1}  \int_{[0,\infty)^{i+1}}f_1(x_{i+1})\nonumber\\
&&\times~\!L_{\hspace{-0.1em}D}\!\left(x_v,x_{\sigma(v)}\right)\prod_{j=1}^{i}\prod_{s\in p_j}f_{s}(x_j)  \prod_{1\leq j \neq v \leq i+1}  K\!\left(x_j, x_{\sigma(j)}\right)\prod_{j=1}^{i+1}\dd x_j,
\end{eqnarray}
while $H_{\rvert P_q\rvert+1}^{(2)}\!\left(X_1,{\bf X}_{P_q}\right)$ and $h_{\rvert P_q\rvert+1}^{(2)}\!\left(X_1,{\bf X}_{P_q}\right)$ are the expression (\ref{eq:dcid2}) with $L_{\hspace{-0.1em}D}\!\left(x_r,x_{\sigma(r)}\right)$ replaced by $L_{\hspace{-0.1em}H}\!\left(x_r,x_{\sigma(r)}\right)$ and $L_{\hspace{-0.1em}h}\!\left(x_r,x_{\sigma(r)}\right)$, respectively. 

Putting together the results (\ref{eq:dcid}), (\ref{eq:dcid2}) and the fact that the summation over partitions in~(\ref{eq:dcIth}) can be divided into two summations, depending on whether the element $1$ is partitioned in a singleton, we arrive at
\begin{eqnarray}
H_{\rvert P_q\rvert+1}^{(1)}\!\left(X_1,{\bf X}_{P_q}\right)+H_{\rvert P_q\rvert+1}^{(2)}\!\left(X_1,{\bf X}_{P_q}\right)&=&H_{\rvert P_q\rvert+1}\!\left(X_1,{\bf X_{P_q}}\right),\label{eq:exitHi2}\\
h_{\rvert P_q\rvert+1}^{(1)}\!\left(X_1,{\bf X}_{P_q}\right)+h_{\rvert P_q\rvert+1}^{(2)}\!\left(X_1,{\bf X}_{P_q}\right)&=&h_{\rvert P_q\rvert+1}\!\left(X_1,{\bf X_{P_q}}\right),\label{eq:exithi2}\\
D_{\rvert P_q\rvert+1}^{(1)}\!\left(X_1,{\bf X}_{P_q}\right)+D_{\rvert P_q\rvert+1}^{(2)}\!\left(X_1,{\bf X}_{P_q}\right)&=&D_{\rvert P_q\rvert+1}\!\left(X_1,{\bf X_{P_q}}\right),\label{eq:exitdi2}
\end{eqnarray}
where $H_{\rvert P_q\rvert}({\bf X}_{P_q})$, $h_{\rvert P_q\rvert}({\bf X}_{P_q})$, and $D_{\rvert P_q\rvert}({\bf X}_{P_q})$ denote the integral
\begin{eqnarray}\label{eq:dcIth}
&&\sum_{\left\{p_1,\ldots,p_i\right\}\in\mathcal{P}_{P_q}}(-1)^{i-1}\sum_{\sigma\in C_{P_q}}\sum_{v=1}^{i}\int_{[0,\infty)^i}\!\!L\!\left(x_v,x_{\sigma(v)}\right)\prod_{j=1}^{i}\prod_{s\in p_{j}}f_s\!\left(x_j\right)\nonumber\\
&&\times\prod_{1\leq j \neq v \leq i} K\!\left(x_j,x_{\sigma(j)}\right)\prod_{j=1}^{i}\dd x_{j}
\end{eqnarray}
with $L(x,y)$ being $L_H(x,y)$ in~(\ref{eq:LH}), $L_h(x,y)$ in~(\ref{eq:Lh}), and $L_D(x,y)$ in~(\ref{eq:LD}), respectively.

Now inserting~(\ref{eq:t1p1a1s}),~(\ref{eq:t1p1a2s}), and~(\ref{eq:t1p1a3s}) into~(\ref{eq:t1p1a123}), the decoupled term $\delta_l(k)$ is expressed as
\begin{eqnarray}\label{eq:th10}
\delta_l(k)&=&\frac{1}{2}\sum_{r=0}^{k-1}\sum_{s=0}^{l-2}\frac{(l-2)!}{s!(l-2-s)!}\Big(\kappa(R)\big(H_{s+1}(X_1,T,\dots,T) h_{l-1-s}(X_l,T,\dots,T)\nonumber\\
&&+~\!h_{s+1}(X_1,T,\dots,T)H_{l-1-s}(X_l,T,\dots,T)\hspace{0.1em}\big) \nonumber \\
&&-~\!2D_{s+1}(X_1,T,\dots,T)D_{l-1-s}(X_l,T,\dots,T)\Big),
\end{eqnarray}
where $X_1$ and $X_l$ are respectively the linear statistics~(\ref{eq:th1f1}) and~(\ref{eq:th1fl}). In~(\ref{eq:th10}), due to the symmetry of statistics in~(\ref{eq:th1ft}), we have written the sum over partitions $\left\{\hspace{-0.1em}P_1, P_2\right\}\in \mathcal{B}$ as a sum over the cardinality, denoted by $s$, of the subset $P_1$ with
\begin{equation}
\binom{l-2}{s}=\frac{(l-2)!}{s!(l-2-s)!}
\end{equation}
counting the symmetric cases for each $s$. By Propositions~\ref{props},~\ref{propdk} and Corollary~\ref{codkb}, the decoupled integrals in (\ref{eq:th10}) are expressed in terms of joint cumulants of order no more than $l-1$. Shifting the summation index $s\to s-1$ in~(\ref{eq:th10}) while keeping in mind the shorthand notations~(\ref{eq:dcs1}) and~(\ref{eq:dcs2}) yields the claimed expression~(\ref{eq:dT}) of decoupled term. 

In computing cumulants $\kappa_l(T)$ using Theorem~\ref{Theorem1}, one also needs to decouple joint cumulants involving ancillary statistics $R_k$ concealed in~$D_{l,s}(k)$ of~(\ref{eq:dT}). Specifically, as seen from the result~(\ref{eq:coro11}), the term 
\begin{equation}\label{eq:Rdt}
-\sum_{r=0}^{k-1}\sum_{s=0}^{l-2}\frac{(l-2)!}{s!(l-2-s)!}D_{s+1}(X_1,T,\dots,T)D_{l-1-s}(X_l,T,\dots,T)  
\end{equation} 
of the decoupled term $\delta_l(k)$ in~(\ref{eq:th10}) consists of joint cumulants up to $\kappa_{l-1}(R_{k-1},T,\dots,T)$, which a new decoupling structure is needed for $k,l\geq3$. The corresponding decoupling structure turns out to be
\begin{eqnarray}\label{eq:th2d}
\!\!\!\!\!\!\kappa_l(R_k,T,\dots,T)-\frac{\dd }{\dd\alpha}\kappa_{l-1}(R_{k+1},T,\dots,T)+\kappa_{l-1}^\prime(T_k,T,\dots,T)=\delta_l^{(R)}(k),
\end{eqnarray}
which is obtained by applying Proposition~\ref{propdab} and Proposition~\ref{propdk} along with the property of joint cumulants involving a constant~\cite{PT11}
\begin{equation}
\kappa(R_{0},{\bf X})=0
\end{equation}
to the definition below, cf.~(\ref{eq:dlk4}),
\begin{eqnarray}
\delta_l^{(R)}(k)&=&\kappa_l(R_k,T,\dots,T)-\kappa_{l}(R_{k+1},T,\dots,T,T_0) \nonumber \\
&&+~\!\kappa_{l}(R,T,\dots,T,T_k)-\kappa_{l}(R_0,T,\dots,T,T_{k+1}). \label{eq:delta2}
\end{eqnarray}
Comparing~(\ref{eq:delta2}) to~(\ref{eq:dlk4}), the recycling of the decoupled term $\delta_l^{(R)}(k)$ is read off from~(\ref{eq:th10}) when considering the specialization  
\begin{eqnarray}
X_1 &=& R_{r}, \\ 
X_l &=& T_{k-1-r}
\end{eqnarray}
as
\begin{equation}\label{eq:dR}
\delta_{l}^{(R)}(k)=\sum_{s=1}^{l-1}\frac{(l-2)!}{(s-1)!(l-s-1)!}\left(\kappa(R)H^{(R)}_{l,s}(k)-2D^{(R)}_{l,s}(k)\right),
\end{equation} 
where
\begin{eqnarray}
H_{l,s}^{(R)}(k) &=& \sum_{r=0}^{k-1}\big(H_s(R_r,T,\dots,T)h_{l-s}(T_{k-r-1},T,\dots,T) \nonumber \\
&&+~\!\hspace{0.1em}h_s(R_r,T,\dots,T)H_{l-s}(T_{k-r-1},T,\dots,T)\big), \label{eq:dcsr1} \\
D_{l,s}^{(R)}(k) &=& \sum_{r=0}^{k-1}D_s(R_r,T,\dots,T)D_{l-s}(T_{k-r-1},T,\dots,T). \label{eq:dcsr2}
\end{eqnarray}
This completes the proof of Theorem~\ref{Theorem1}.
\end{proof}

\section{Proof of Corollary \ref{coro2}}\label{appa:coro2} 
\begin{proof}
The first task of the proof is to show that the mean value of $T_k$ for any positive integer $k$ is of the form
\begin{equation}\label{eq:tl}
\kappa(T_k)=a_k\psi_0(m+\alpha)+\text{NP},
\end{equation}
where the coefficient $a_k$ will be shown to be the mean of $R_k$,
\begin{equation}\label{eq:al}
a_k=\kappa\!\left(R_k\right).
\end{equation}
Here, NP denotes non-polygamma terms that are in fact polynomials in $m$ and $\alpha$, where NP may be different for each use. 

To show~(\ref{eq:tl}), it is convenient to introduce another recurrence relation of $\kappa(R_k)$ shown in~(\ref{eq:rrecurL}) as proved below. We start by writing the recurrence relation~(\ref{eq:rrecur}) valid for $k\in\mathbb{R}_{\ge0}$ as
\begin{equation}\label{eq:rrecur1}
(k+1)\kappa\!\left(R_k\right)=(k-1)(2m+\alpha)\kappa\!\left(R_{k-1}\right)+m(m+\alpha)\left(H(R_{k-1})+h(R_{k-1})\right),
\end{equation}
where the definitions
\begin{eqnarray}
H(R_{k}) &=& \kappa^{\normalfont\plus}\!\left(R_{k}\right)-\kappa\!\left(R_{k}\right) = \frac{m!}{(m+\alpha)!}\int_0^{\infty}x^{k}w(x)L_{m}^{(\alpha)}(x)L_{m}^{(\alpha)}(x)~\!\dd x, \\
h(R_{k}) &=& H(R_{k})\rvert_{m\to m-1}=\kappa\!\left(R_k\right)-\kappa^{\normalfont\minus}\!\left(R_{k}\right)
\end{eqnarray}
are in line with those in Table~\ref{table2}. With the help of recurrence relation~(\ref{eq:recur}) and structure relation~(\ref{eq:Lsr}), integration by parts of 
\begin{equation}
\int_{0}^{\infty}\!\left(\frac{\dd}{\dd x}x^{k}\right)w(x)L_{m-1}^{(\alpha)}(x)L_{m}^{(\alpha)}(x)~\!\dd x
\end{equation} 
and
\begin{equation}
\int_{0}^{\infty}\!\left(\frac{\dd}{\dd x}x^{k}w(x)\right)L_{m}^{(\alpha)}(x)L_{m}^{(\alpha)}(x)~\!\dd x
\end{equation}
leads to the identities 
\begin{equation}\label{eq:rmm-1m}
m(m+\alpha)(H(R_{k-1})-h(R_{k-1})) = k(k-1)\kappa\!\left(R_{k-1}\right)
\end{equation}
and
\begin{equation}\label{eq:Hk-1k-2} 
H(R_{k-1})-\left(2m+\alpha-1+k\right)H(R_{k-2})=2(k-2)\kappa(R_{k-2}),
\end{equation}
respectively. Eliminating $H(R_{k-1})$ and $h(R_{k-1})$ from~(\ref{eq:rrecur1}),~(\ref{eq:rmm-1m}), and~(\ref{eq:Hk-1k-2}), we arrive at the desired recurrence relation
\begin{eqnarray}\label{eq:rrecurL}
\!\!\!\!\!\!\!\!\!(k+1)\kappa\!\left(R_k\right)=(2k-1)(2m+\alpha)\kappa(R_{k-1})+(k-2)\left((k-1)^2-\alpha^2\right)\kappa(R_{k-2}).
\end{eqnarray}
Note that the result~(\ref{eq:rrecurL}), valid for nonnegative real $k$, was established in~\cite{HT03} under the assumption of positive integer $k$. Taking derivative of (\ref{eq:rrecurL}) with respect to $k$ yields
\begin{eqnarray}\label{eq:recurtk2}
\!\!\!\!\!\!\!\!\!\!\!\!\!(k+1)\kappa\!\left(T_{k}\right)&=&(2k-1)(2m+\alpha)\kappa\!\left(T_{k-1}\right)+(k-2)\left((k-1)^2-\alpha^2\right)\kappa\!\left(T_{k-2}\right) \nonumber \\
&&+~\!\text{NP},
\end{eqnarray}
where NP collects terms involving $\kappa\!\left(R_{k}\right)$ that, according to~(\ref{eq:rrecur}), are non-polygamma. The claimed form~(\ref{eq:tl}) is now seen inductively by inserting into~(\ref{eq:recurtk2}) the initial values, obtained from Lemma~\ref{lemmart},
\begin{eqnarray}
\kappa\!\left(T\right)&=&a_1\psi_0(m+\alpha)+\frac{1}{2}m(m+1), \\
\kappa\!\left(T_2\right)&=&a_2\psi_0(m+\alpha)+\frac{m}{6}\left(10m^{2}+9m\alpha+6m+3\alpha+2\right)
\end{eqnarray}
with
\begin{eqnarray}
a_1&=&\kappa\!\left(R\right)=m(m+\alpha),\label{eq:ar1}\\
a_2&=&\kappa\!\left(R_2\right)=m(m+\alpha)(2m+\alpha).\label{eq:ar2}
\end{eqnarray}
Inserting~(\ref{eq:tl}) into~(\ref{eq:recurtk2}), by matching the coefficients of the digamma function $\psi_0(m+\alpha)$ on both side of the equation, one obtains
\begin{equation}\label{eq:rrecura}
(k+1)a_k=(2k-1)(2m+\alpha)a_{k-1}+(k-2)\left((k-1)^2-\alpha^2\right)a_{k-2}.
\end{equation}
The above recurrence relation of $a_k$ is the same as that of $\kappa(R_k)$ in~(\ref{eq:rrecurL}). This establishes~(\ref{eq:al}).

The remaining task of the proof is to show that the coefficient $a_l=\kappa\!\left(R_l\right)$ of the digamma function $\psi_0(m+\alpha)$ in the mean expression $\kappa(T_l)$ carries over, through the execution of Theorem~\ref{Theorem1}, to the coefficient of the highest-order polygamma function $\psi_{l-1}(m+\alpha)$ in the $l$-th cumulant expression $\kappa_l(T)$. This is seen from Algorithm~\ref{alg} that implements the decoupling procedure of Theorem~\ref{Theorem1}, where, in each of the $l-1$ steps from the starting point $\kappa(T_l)$ to the final expression $\kappa_l(T)$, the corresponding highest-order polygamma function can only be generated from the decoupling statistics 
\begin{equation}
\frac{\dd}{\dd\alpha}\kappa_{L-1}\left(T_{l-L+2},T,\dots,T\right)     
\end{equation}
but not from the decoupled terms
\begin{equation}
\delta_L(l-L+1)
\end{equation}
for $L=2,\dots,l$. Therefore, the polygamma function of the highest order in $\kappa_l(T)$, hence that in $\mathbb{E}\!\left[T^l\right]$ due to the cumulants to moments relation (\ref{eq:mk}), is
\begin{equation}\label{eq:Thp}
\kappa(R_l)\psi_{l-1}(n)
\end{equation}
when keeping in mind the definitions~(\ref{eq:alpha}) and~(\ref{eq:polygamma}). Inserting~(\ref{eq:Thp}) into the moment conversion formula~(\ref{eq:TSC}) between $S$ and $T$, the terms that contribute to the highest-order polygamma functions in $\mathbb{E}\!\left[S^l\right]$ are
\begin{equation}\label{eq:c2i}
(-1)^{l}\frac{\Gamma(mn)\kappa(R_l)}{\Gamma(mn+l)}\psi_{l-1}(n)+A_0,
\end{equation}
where the highest-order polygamma function in $A_0$ is computed by the definition of Bell polynomials~(\ref{eq:cbp2}) as 
\begin{equation}\label{eq:c2psi}
(-1)^{l-1}\psi_{l-1}(mn+l).  
\end{equation}
Employing the definition of Pochhammer's symbol
\begin{equation}
\frac{\Gamma(mn+l)}{\Gamma(mn)}=(mn)_{l}
\end{equation}
in~(\ref{eq:c2i}) before shifting the argument of polygamma function in~(\ref{eq:c2psi}) by using~(\ref{eq:psisre}) completes the proof of Corollary~\ref{coro2}.
\end{proof}

\end{document}